\documentclass[twocolumn,showpacs,preprintnumbers,amsmath,amssymb,nofootinbib]{revtex4-1}

\usepackage{graphicx}
\usepackage{dcolumn}
\usepackage{bm}
\usepackage{verbatim} 
\usepackage{epsfig}
\usepackage{amsmath}
\usepackage{amsthm}
\usepackage{subfigure}

\usepackage{hyperref}

\begin{document}

\date{2 February, 2023}

\title{Proposed experimental study of wave-particle duality in $p,p$ scattering}
\author{Richard M Talman}
\affiliation{
Laboratory for Elementary Particle Physics,
Cornell University, Ithaca, NY, USA}

\begin{abstract}
Of all nuclear physics experiments none are more fundamental than ``elastic'' $p,p$ and, secondarily, 
$p,d$ or $d,d$ scattering.  Recognizing that these particles are themselves composite, ``elastic'' 
scattering may be accompanied by temporary internal rearrangement with undetectably small energy loss.  
This paper argues initially that correct calculation of the spin dependence of $p,p$ 
(and other charged particle) elastic scattering must account for a previously-neglected relativistic effect 
of ``$G$'', the anomalous magnetic dipole moment (MDM) of the scattering particles.  The paper then precedes 
to describe storage ring $p,p$ scattering configurations capable of confirming this contention. 
Especially important experimentally for protons is the existence of ``perfect'' (greater than 99\%) 
proton-carbon scattering polarimetric analyzing power $A$ at $K=183.1$\,MeV laboratory kinetic energy 
and nearly as high nearby. 
Possibilities:
(i) In a storage ring collider with counter-circulating proton beams, each with, say, 
$K=200\,$MeV energy, the final spin states of coincident scattered protons can be determined 
with high probability for a significantly large fraction of all scatters, both prompt and delayed. 
For comparison with current descriptions based on proton scattering from a hydrogen target fixed in the 
laboratory, this corresponds roughly, to proton kinetic energy $K=400\,$MeV in the laboratory frame,
close to the pion production threshold.
(ii) In a ``DERBENEV-style'' figure-8 storage ring, independently polarized, diametrically opposite bunches
on orthogonal orbits can collide at the beam crossover point with symmetric $K''\approx200$ MeV energies 
in a slow, transversely moving frame.
(iii) As another compromise, $p$ and $d$ beams can counter-circulate at the same time in a small racetrack 
shaped ring with superimposed electric and magnetic bending. In this case the scattering would be ``WOLFENSTEIN-style'', 
with collinear incident orbits (at the cost of significantly inferior polarimetry for the deuteron beam).
To investigate the consistency of quantum mechanics and special relativity it is proposed to implement 
options (ii) and (iii) in the COSY beam hall.

\end{abstract}


\maketitle

\tableofcontents
\setcounter{tocdepth}{2}

\section{Introduction} 
\emph{This paper is a companion to two published papers with related subject matter\ \cite{RT-ICFA}\cite{RT-CLIP-PTR},
and a planned paper on storage ring studies of nuclear transmutation.}

As everybody knows, for laboratory kinetic energies up to approximately 100\,MeV (at which point the proton deBroglie
wavelength is somewhat greater than the proton diameter) the quantum (QM) and classical (CM) (i.e. Rutherford) descriptions of 
elastic $p,p$ scattering agree splendidly. See Figure~\ref{fig:deBroglie-lambda-vs-E_tot}. 
This agreement clearly could not and cannot continue beyond, 
say, 400\,MeV, when pions, unknown to Rutherford, began or begin to appear.  See Fig.~\ref{fig:pp-total-crsctn-vs-KE}.

This failure with increasing energy has to be ascribed to one of three sources: previous ignorance of spin, of new 
particles such as pions, or of wave mechanics (WM) (i.e. wave/particle duality). Since the failure is sudden, while the 
wavelength variation is smooth, one has to blame the discontinuous behavior either on spin or on new particle threshold, 
neither of which Rutherford had anticipated.  

For present purposes this justifies modest extrapolation of classical Rutherford treatment to higher energy, 
while continuing to treat ``particles'' as particles, rather than as waves, at least for purposes of approximate 
description.  As regards initial beam preparation, this continues to be standard practice at all higher proton energies, 
even though it is misleading in some contexts, especially for charged particles such as the electron and, 
arguably, the muon.   

One conjectures then, that semi-classical, Rutherford style Newtonian classical mechanics (CM) description can 
be trusted to some modest extent, for phenomenological extrapolation toward higher energy, by the ad hoc empirical
inclusion of new particle thresholds as they becomes necessary.

An intuitively attractive way of merging these conflicting theoretical pictures is to segregate scattering events 
temporally into ``prompt'' and ``delayed'' categories, with the expectation that the prompt events are well described by 
both CM and QM, while ``delayed'' events require QM.  Rephrasing this, one imagines ``prompt events'', for which
classical physics provides definitive (but eventually incorrect) results, and ``delayed events'' for which
quantum mechanics QM is required. QM (i.e. wave mechanics (WM)  at quantum, nuclear or atomic length scale) has the more 
challenging task  of providing probabilistic ``long term'' description of nuclear scattering, elastic plus inelastic.
Resonance, Heisenberg time uncertainty, and barrier penetration go hand in hand during the delayed phase.  

In present context, it has already been shown experimentally that QM subsumes (beyond a reasonable doubt) the 
correct classical description of low energy Rutherford scattering.  What remains to be investigated experimentally 
is the degree to which the segregation into ``prompt'' and ``delayed'' categories is useful. 

One anticipates few surprises.  Nuclear physics theory already anticipates such distinct time scales.
One can scarcely improve upon the abstract of reference\ \cite{Gomez-time-scales}
\noindent
``Two relevant time scales are introduced to describe the interplay of nuclear structure 
and nuclear reactions for exotic nuclei. The collision time represents the time 
dependence of the external field created by the target on the projectile. The 
excitation time represents the characteristic time dependence of the projectile 
degrees of freedom due to its internal Hamiltonian. The comparison of these two 
time scales indicate when approximate treatments of the reaction, such as the sudden 
approximation, implicit in the eikonal treatment, is applicable.''
\footnote{An example of ``eikonal treatment'' is provided by Fig.~\ref{fig:EikonalTrajectories}.}

For prompt behavior, one expects agreement with high probability between deterministic classical treatment and 
probabilistic quantum mechanical treatment. On longer time scales, especially in the light of barrier penetration
by ``tunneling'' and new particle production, one can only expect theory to produce a probabilistic description 
of the subsequent evolution.

This paper concentrates on the detection, in a colliding beam storage ring, of coincident scattered protons 
coming to rest in graphite polarimeter chambers providing nearly full directional coverage.  This makes it 
practical to test the quantum mechanical description of low energy nuclear physics with unprecedented sensitivity. 

The emphasis will be more on spin dependence than on time dependence, but curious event delays as great as
10\,ns may be detectable.

\section{Goals of the paper\label{sec:Goals}}
\subsection{Rationale\label{sec:Rationale}}
Since its inception, between the first two world wars nuclear physics has gradually and naturally split into
related but largely disjoint fields: low energy-, high energy- (or elementary particle-), astro-, cosmological-,
quantum information-, and so on. There has been spectacular (but uneven) success in every one of these areas.  

By now a certain natural sense of self-satisfaction, but also resignation, has developed.  This is manifested, 
for example, by the phrase ``physics beyond the standard model''.  This phrase implied initially that 
``a correct foundation has been laid'', but has come, after many years to include ``but there seems to be nothing 
further that can be done to check this, at least at current funding levels''.

The attitude expressed in this paper is exactly opposite; it is that ``a correct foundation has not been laid,
and there is plenty that can be done about it at present funding levels''.  The 2022 Nobel prize “for experiments 
with entangled photons, establishing the violation of Bell inequalities and pioneering quantum information science”
provides an example of success using this approach; especially to the extent that entanglement is interpreted 
as a fundamental defect, rather than a natural refinement in the interpretation of quantum mechanics and field theory.

This paper concentrates on testing experimentally the quantum mechanical treatment of low energy nuclear physics, 
with special concentration on the role of nuclear spin in elastic scattering.  
This includes, for example, questioning the standard treatment of identical particles and investigating the possibility 
that the entanglement of scattered protons be detected.  Special attention is also paid, 
possibly for the first time, to the influence of the anomalous magnetic moment in $p,p$ scattering; noting, as an
aside, that the adjective ``anomalous'' means ``not understood''.  See Appendix~\ref{sec:MDMs}.

For more than half a century the single most mysterious aspect of elementary particle physics has concerned the 
mass of the muon ($\mu$-meson), a particle discovered by Anderson and Neddermeyer at Caltech in 1936.  The muon 
spin was measured to be 1/2 in 1960 by Hughes and others .  Expressed as rest energy, the $\mu$-mass is 105.66 MeV 
(accepting that a particle with half-life of 2\,$\mu$s can have an unambiguous mass). 

Already mysterious when spin was introduced was the fact that a measurably point-like particle, the electron, 
could have  non-zero angular momentum and magnetic dipole moment.  For a charged particle with 200 times greater 
mass, yet angular momentum corresponding to that of a heavy electron, seems to imply that the muon, too, is a 
point particle (lepton).  Implying infinite density, this calls into question our understanding of ``mass''. 
\footnote{Reading just the thirteen pages of Chapter~2 
of the Max Jammer book ``Concepts of Mass'', one may be surprised to learn that, as recently as the time of Archimedes,
the Greeks had a correct understanding of \emph{specific gravity} while, at the same time, no concept of 
\emph{density}.  With no concept of mass, there can be no way to conceive of density.  For us, except for being 
expressed in different dimensional units, we consider specific gravity and density to be essentially identical 
material properties.}
   
The paper is, however, primarily experimental in nature; describing in  detail, accelerator designs 
capable of detecting possible inconsistencies in the current understanding of low energy nuclear physics,
with special attention to time reversal invariance and the role of anomalous magnetic moments.

\subsection{Glossary of ring acronyms and properties}
In this paper there are enough different accelerators, with unfamiliar acronyms, to justify introducing them 
now, in one place.  COSY is a low energy proton or deuteron storage ring in Juelich, Germany, whose days 
are numbered.  Reference\ \cite{CYR} is a long and detailed CERN yellow report, slightly pre-dating COVID, 
which describes PTR as a prototype ring proposed to be built in the COSY beam hall, along with a bunch 
accumulator BA.  By retaining existing equipment, especially injection, already functioning in COSY, 
a substantial fraction (perhaps 1/3) of the value of the next generation installation in the COSY hall
would be saved.  Figures in this paper show the COSY beam hall with arc elements in their current 
locations, but with altered straight sections.  
  
Two (mutually compatible) storage ring designs are described, which have been produced during the 
COVID ``hiatus''.  One is shaped like a race-track, the other like a figure-8. Though these rings cannot 
operate at the same time; to switch from one to the other operation can be expected to be routine.
Together, they are more powerful than had previously beam contemplated.  Furthermore, these designs 
retain a much larger fraction of the current COSY installation; as already mentioned, the circular arcs 
of the COSY ring are retained in plane, common to both storage rings, and injection can be retained as-is.
Yet 2/3 of the value invested in COSY in the past is now retained.

\begin{figure*}[hbt]
\centering
\includegraphics[scale=0.3]{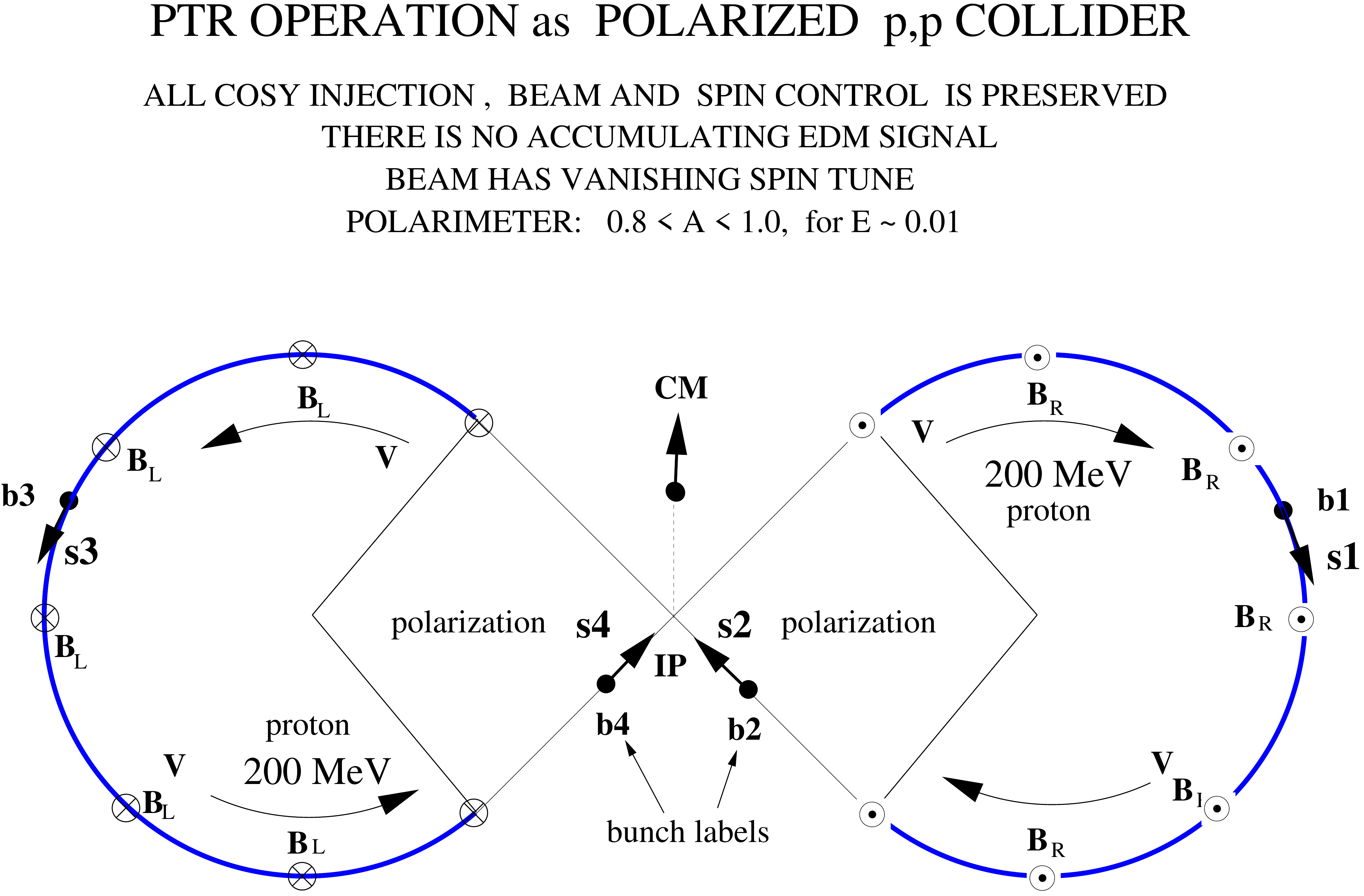}
\caption{\label{fig:Equal-period-pp-collider-mod-mod}Skeleton design plan for a 200\,MeV (kinetic, laboratory) energy 
polarized $p,p$ figure-eight collider.  Since the global spin tune vanishes ($Q_s=0$) bunch polarizations can be set independently 
and can be phase locked (shown pointing forward as {\bf s2} and {\bf s4} for bunches 
{\bf b2} and {\bf b4}) to remain frozen indefinitely.  
 \emph{Two beams cannot counter-circulate at the same time}.  But beam circulation direction can be 
reversed with frequency domain setting and re-setting precision, without the need for (impractically precise) 
electric and magnetic field measurement.  This is almost as useful as simultaneous counter-circulation for reducing
significant systematic errors by averaging over reversals.
Both incident beam spins states at the collision intersection point (IP) can be pure, and each scattered
particle momentum and polarization are measured with maximum possible (100\%) analyzing power as each it
slows through 183\,MeV energy.}
\end{figure*}

\begin{figure}[hbt]
\centering
\includegraphics[scale=0.26]{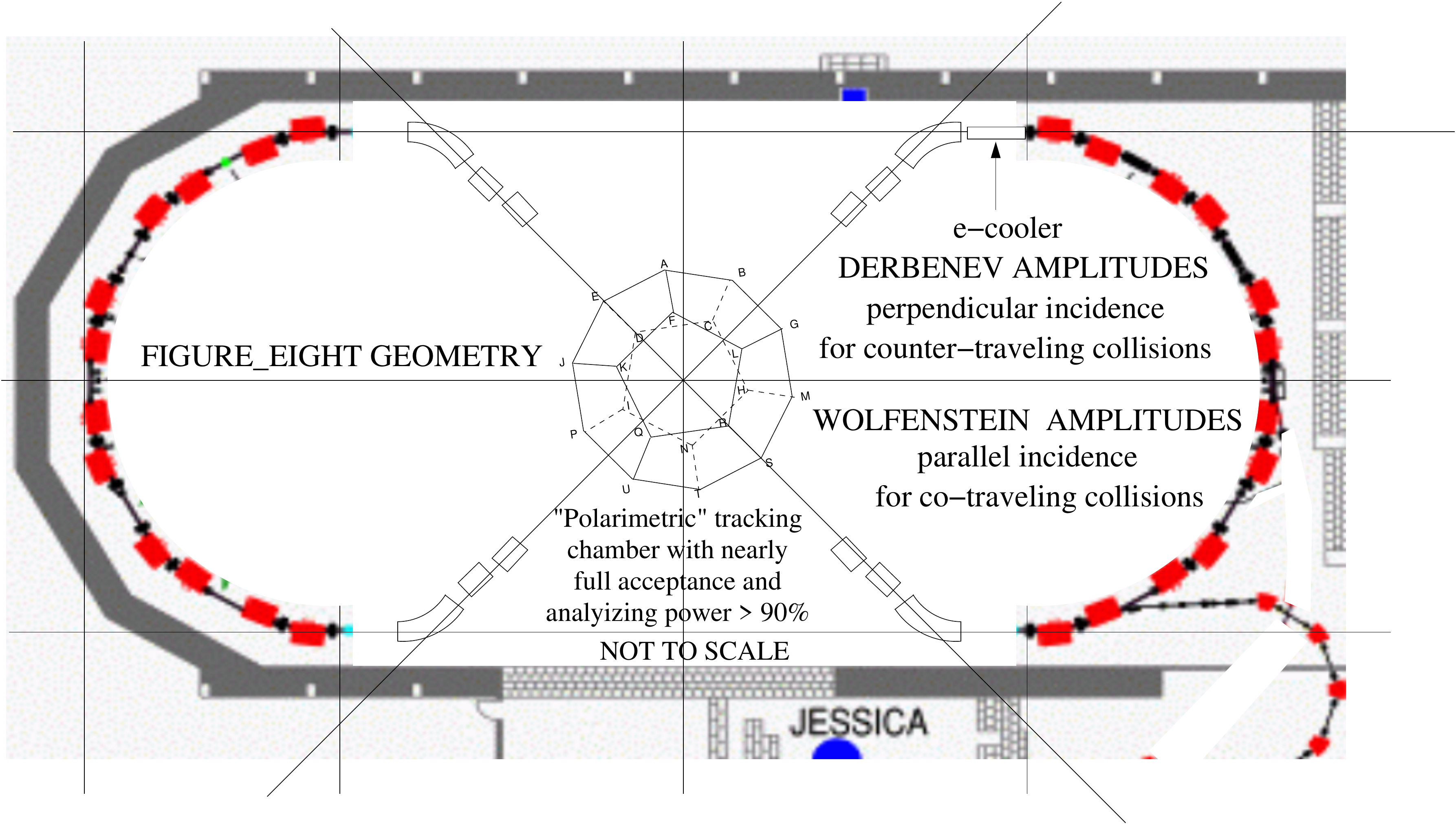}
\caption{\label{fig:CPT-FigureEight-layout}This layout figure shows COSY (as of 2022)
after incorporation of the superimposed electric bending shown in Figure~\ref{fig:EM-superimposed-COsY-racetrack}. 
The semicircular arcs are preserved exactly as at present, except the arc magnets are powered individually, to
compensate appropriately for the superimposed electric bending.  ``
Co-traveling bunches'' 
(needed for resonant nuclear transmutation) have same sign, parallel incident momenta. ``Counter-traveling bunches'' 
(needed for $p,p$ scattering) also have parallel incidence momenta.   The detection chamber situated at the crossing point 
is shown on the right side of Fig.~\ref{fig:COSY-hall-mod6-racetrack}
}
\end{figure}

Figure~\ref{fig:Equal-period-pp-collider-mod-mod} provides a skeleton view of a figure-eight storage ring for
which a more detailed design is shown in Figure~\ref{fig:CPT-FigureEight-layout}.
We designate this configuration as DERBENEV who introduced figure-8 design based on spin consideration.
Most significant of this configuration for present purposes is the fact that a single beam can``collide 
with itself'' or, rather, that bunches separated by half the circumference collide at the crossing 
point,  with orthogonal incident orbits. \emph{This feature of {\rm DERBENEV} configuration contrasts with all previous 
scattering configurations for which colliding bunch orbits are collinear.}

\begin{figure*}[hbt]
\centering
\includegraphics[scale=0.32]{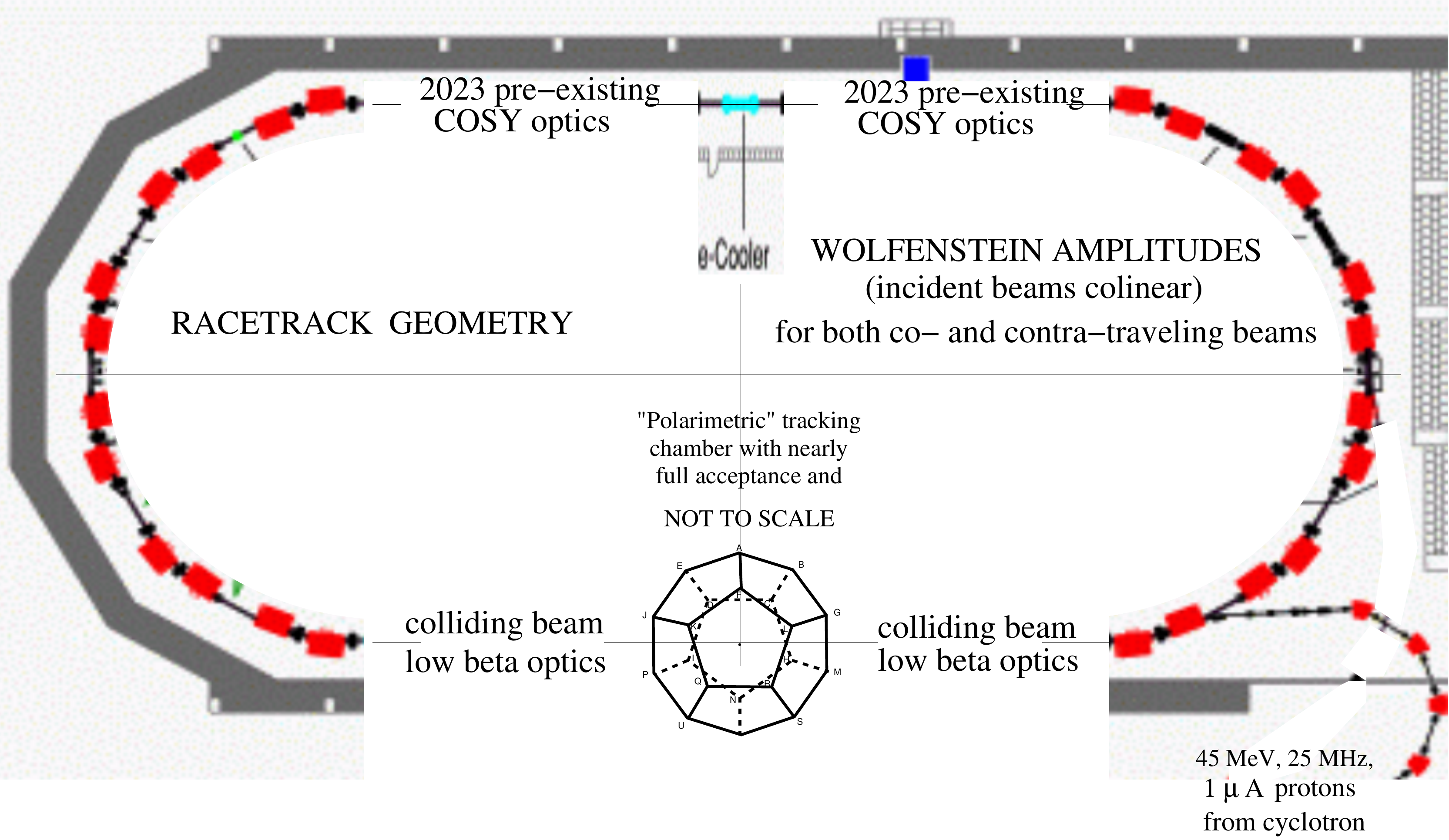}
\includegraphics[scale=0.22]{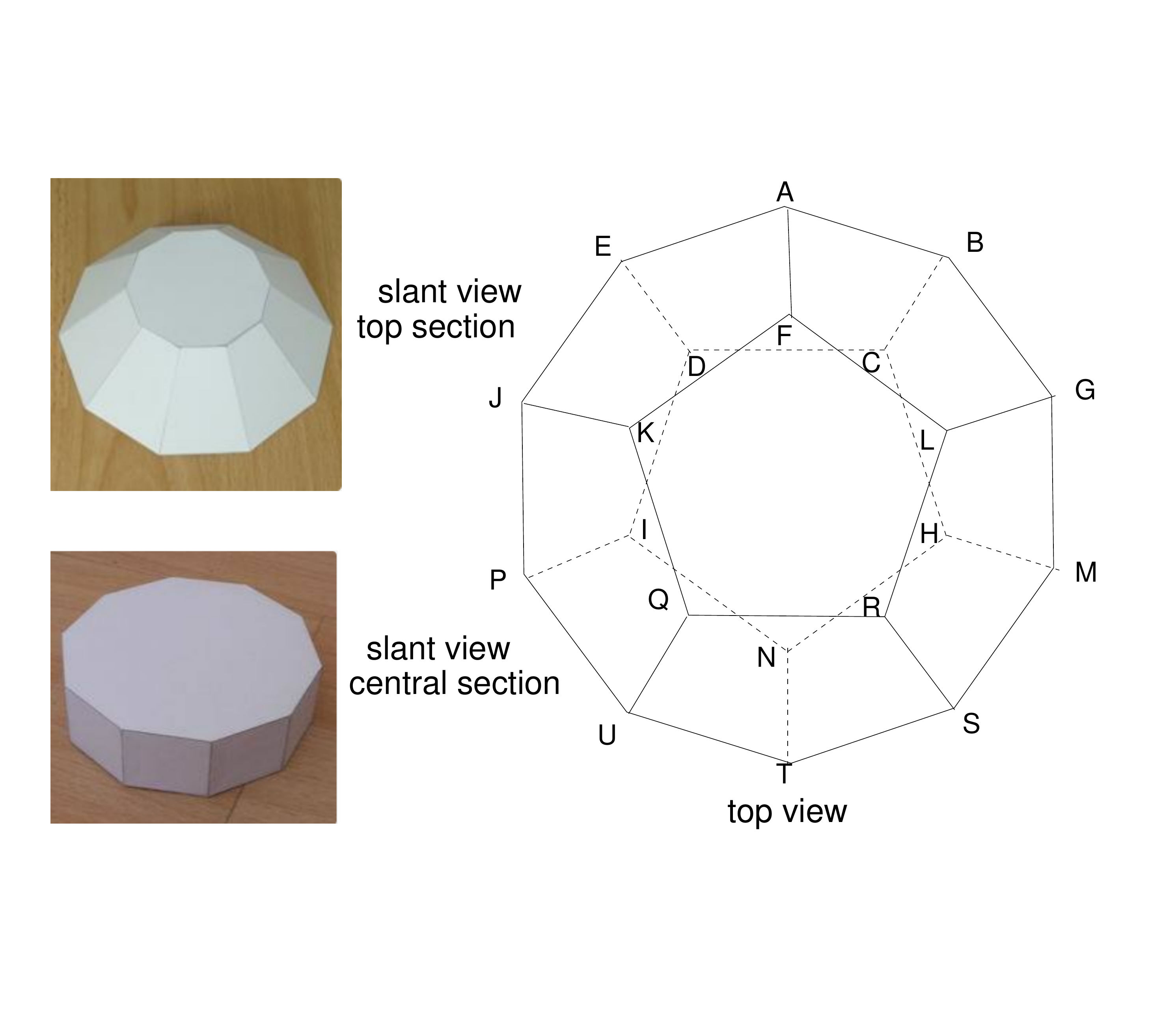}
\caption{\label{fig:COSY-hall-mod6-racetrack}This layout figure shows COSY (as of 2022)
after incorporation of the superimposed electric bending shown in Figure~\ref{fig:EM-superimposed-COsY-racetrack}. 
The semicircular arcs are preserved exactly as at present, except the arc magnets are powered individually, to
compensate appropriately for the superimposed electric bending.  ``
Co-traveling bunches'' 
(needed for resonant nuclear transmutation) have same sign, parallel incident momenta. ``Counter-traveling bunches'' 
(needed for $p,p$ scattering) also have parallel incidence momenta.  
{\bf Right:\ }Artist's conceptions of almost full-acceptance tracking/stopping/polarimeter chambers 
at the intersection point in the lower straight section.  The top and bottom sections resemble partial Platonic 
dodecahedrons, with 12 identical planar faces, each subtending the same solid angle. Labeled vertices are 
for convenient reference here. For example, a ``clean'' stopping proton passing just below point F
in the top pentagon can be expected to be in coincidence with a clean stopping proton midway between points C and D
in the bottom pentagon.  To accommodate passage of the colliding beams there can only be reduced particle detection in 
the central section.
}
\end{figure*}

\begin{figure*}[hbt]
\centering
\includegraphics[scale=0.6]{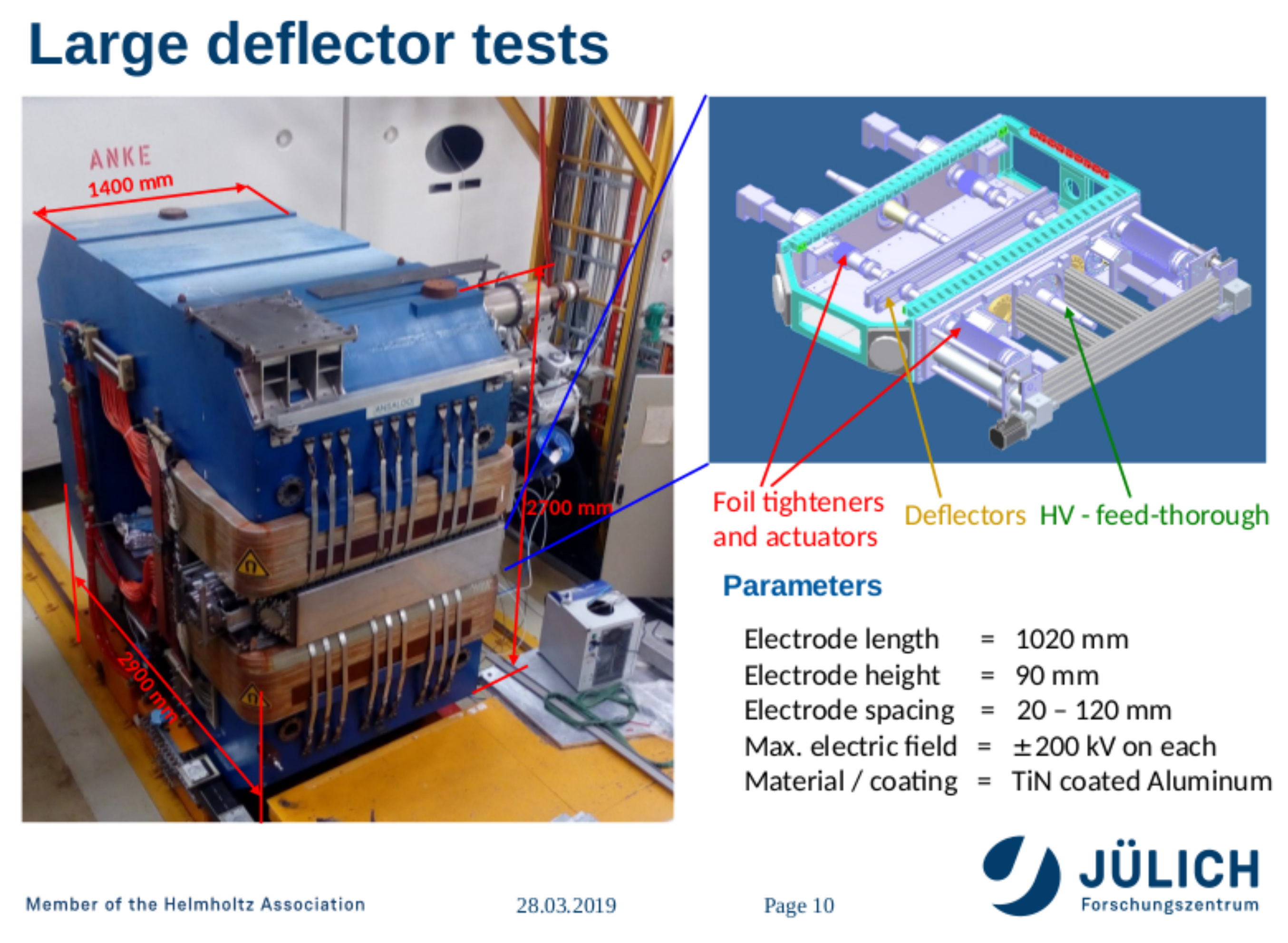}
\caption{\label{fig:EM-superimposed-COsY-racetrack}Grigoryev prototype electric field insert in COSY bending magnets  
for 200\,MeV (kinetic, laboratory) energy operation of COSY with electric bending superimposed on existing magnetic bending.}
\end{figure*}

A second ring design, referred to as WOLFENSTEIN racetrack, is shown in 
Figure~\ref{fig:COSY-hall-mod6-racetrack}.  Simultaneously counter-circulating beams (for example clockwise 
protons and counter-clockwise deuterons) are supported.  This requires electric bending superimposed on the existing
magnetic bending, as illustrated in Figure~\ref{fig:EM-superimposed-COsY-racetrack}. 

\begin{figure*}[hbt]
\centering
\includegraphics[scale=0.45]{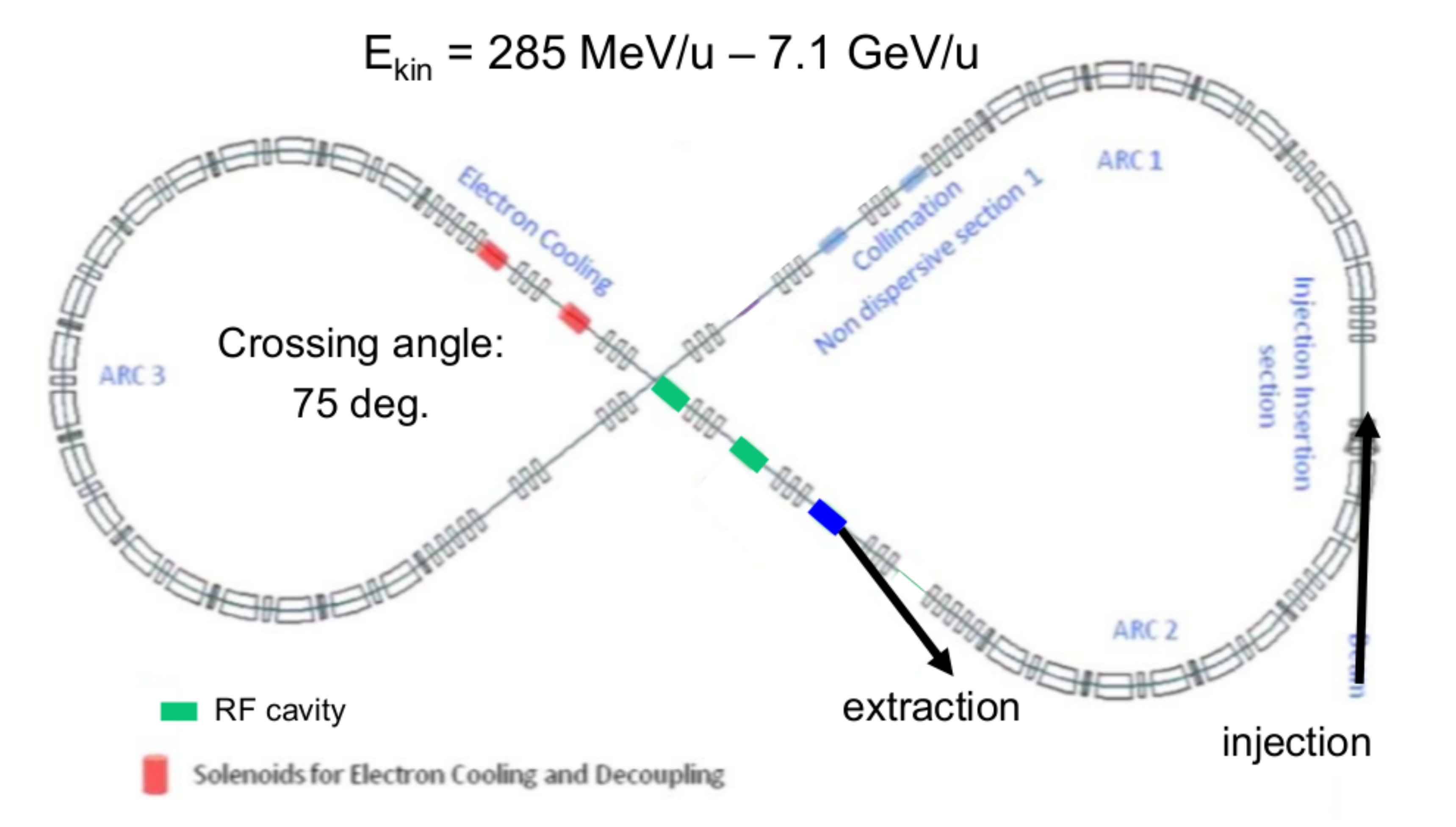}
\includegraphics[scale=0.50]{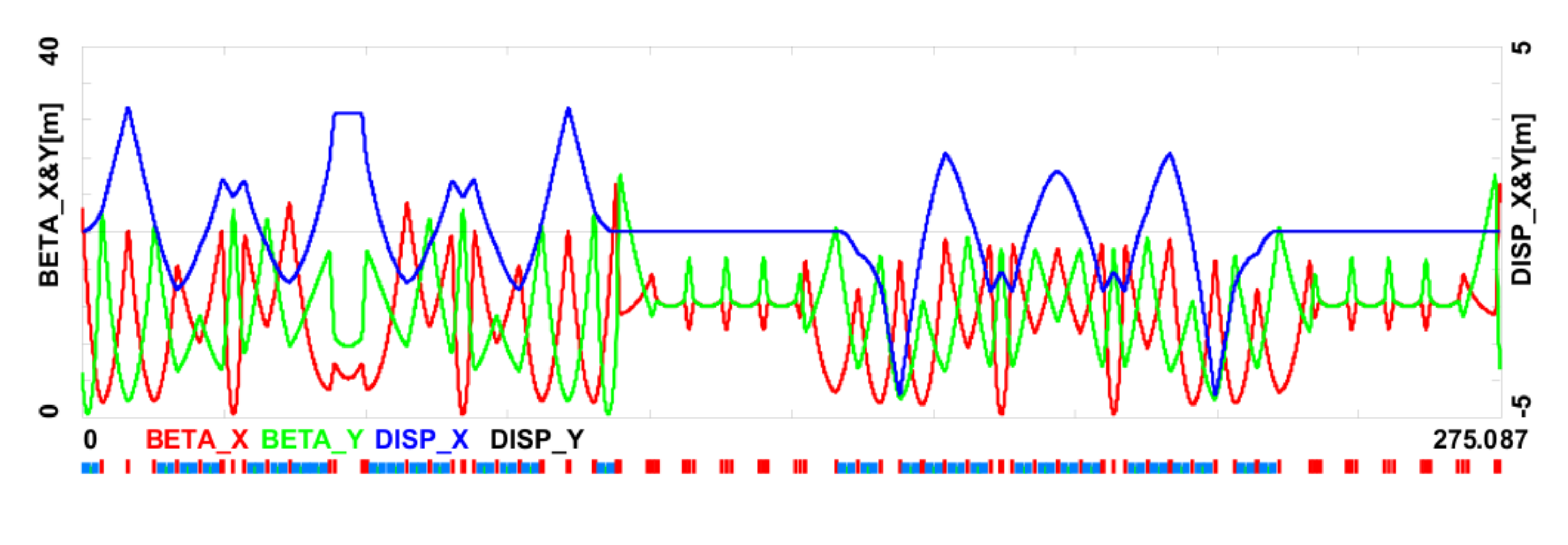}
\caption{\label{fig:JLEIC-booster} These figures apply to the JLEIC-BOOSTER injector accelerator in the Jefforson Lab 
electron-ion collider, JLEIC injector line. Since this figure-eight ring has roughly the same size as the previously 
introduced COSY figure-8 ring, and serves compable beam energies, the 
ring elements and lattice optics can be expected  to be similar. }
\end{figure*}

Yet another figure-eight design is shown, at the top of  Figure~\ref{fig:JLEIC-booster}.  It is the 
JLEIC-BOOSTER injector accelerator as had been intended for the Jefferson Lab electron-ion collider.
At the bottom of the same figure is a graph showing JLEIC-BOOSTER lattice functions.  These figures apply to the 
BOOSTER component in the Jefferson Lab electron-ion collider, JLEIC injector line.  Since this figure-eight ring 
has roughly the same size as the previously introduced COSY hall rings, and serves comparable beam energies, the 
ring elements and lattice optics can be expected to be similar to the lattice functions shown.

Also shown, in Figure~\ref{fig:EM-superimposed-COsY-racetrack}, is an illustration of an insertion constructed 
by Grigoriev and others, whose purpose is to superimpose electric bending within every existing COSY bending magnet.
It is this electric/magnetic superposition which will enable the simultaneous co- and counter-circulation of
stored beams.  This capability is required to implement item~(iii) of the abstract.


\section{The role of spin in low energy nuclear physics\label{sec:RoleOfSpin}}
\subsection{Spin tune in magnetic/electric storage rings}
With novel experimental treatment of spin being emphasized, it is useful to review properties of spin as they are 
understood in accelerator physics.  All that is needed is the concept of spin tune ``$Q_S$'' (where the 
unconventional subscript ``S'', to be replaced immediately by ``E'' or ``M'',
here serves only to  disambiguate from ``charge'') of a (topologically circular) 
storage ring.  The ``topologically circular'' generalization here means (much as in Amp\`ere's circuital law) that 
the ring can be elliptic or racetrack-shaped, or any other closed, singly-connected planar (2D) 
shape, without changing the numerical value of spin tune $Q_S$. 

For a so-called ``Dirac'' particle in a purely magnetic field, with spin directed, say, in the horizontal forward 
direction like a headlight, the spin remains always horizontal and forward.  For any particular particle type, $Q_M$, 
(which can have either sign) is a measure (as a fraction of $2\pi$ of the spin's horizontal angular deviation, relative 
to the particle  direction, after one complete revolution.  For a non-relativistic Dirac particle 
$Q_M=0$.  Irrespective of the value of $Q_M$, the vertical component of the spin of a particle moving in 2D is conserved. 

Eq.~(3.7) in reference~\cite{RT-CLIP-PTR} provides formulas for 
spin tunes in purely electric and purely magnetic rings as
\begin{equation}
Q_E = G\gamma - \frac{G+1}{\gamma},
\quad
Q_M = G\gamma,
\label{eq:BendFrac.7}
\end{equation} 
where $\gamma$ is the usual relativistic factor and $G$ is the ``anomalous MDM''.  

Values of $G$ for low mass nuclear isotopes are given in Appendix~\ref{sec:MDMs}. This appendix addresses several incidental issues,
including the following:
\begin{enumerate}
\item
Explanation of the relationship between ``magnetic g-factor'', $g$, and ``anomalous MDM'', $G$;
\item
Explaining why both $g$ and $G$ can be usefully approximated by rational ratios $n/d$ of positive integers $n$ and $d$,
where $d$ is not allowed to exceed some large value, 100, for example, chosen to be consistent with the precision 
with which the isotope mass is known and, for unstable isotopes, the number of turns in a storage ring that the isotope
can be expected to survive. 
\item
The rational fraction representations of $G$ provide the fraction of $2\pi$ precession caused by anomalous MDMs. Because
of the motion-induced rest frame magnetic field caused, for example, by elastic scattering of one proton by another, this
precession causes spin evolution, proportional to scattering angle, whose presence would otherwise be interpreted as 
violation of time reversal invariance.
\item
Since the evaluation of $g$ and $G$ depend on isotope mass it is important for the same exact mass to be used in their 
separate evaluations and, furthermore that they be represented by dimensionless ratios (such as mass number $A$, or
a nearby value incorporating the mass defect..
\item
Furthermore, for high precision determinations it is essential to use ``integer arithmetic'' (meaning no decimal points) 
in the application of the conversion formula $(G=g\times m/Z-2)/2)$, thereby preserving rational relationships.      
\end{enumerate}
Though one thinks intuitively, of particle MDMs as parameterizing precession in a magnetic field,
in the rest frame of a charged particle moving at any non-vanishing velocity in a transverse electric field there 
is a non vanishing magnetic field, which causes ``in plane'' precession in the plane perpendicular to the magnetic field.  
This is the source of non-vanishing spin precession of an electrically charged
particle moving in a transverse laboratory electric field.\footnote{Though it is rarely (if ever) noted, the precession of a 
charged particle spin in a transverse electric field is, in some sense, ``dual'' to the Stern-Gerlach deflection of the orbit 
of a charged particle moving in a (non-uniform) magnetic field.  } 

In the current (only weakly relativistic) context, as $\gamma\rightarrow1$, these tunes 
can be approximated by their limiting values;
\begin{equation}
\overset{1\leftarrow\gamma}{Q_E = -1}, \quad \overset{\quad1\leftarrow\gamma}{Q_M = G}.
\label{eq:NR}
\end{equation} 
At high energies, as $\gamma\rightarrow\infty$, $Q_E=Q_M=G\gamma$. 

Note that the sign of $Q_M$ is the same as the sign of $G$, which is positive for
protons---proton spins precess more rapidly than their momenta in magnetic fields. 
Deuteron spins, with $G$ negative, lag their momenta in magnetic fields. 

In an electric field (such as the field of a nearby charged electrode or particle) with $G_E$ positive, 
$Q_E$ increases from -1 at zero velocity, eventually switching sign at the ``magic'' velocity where 
the spins in an all-electric ring are ``frozen'' (relative to the beam direction).  

When a particle spin has precessed through 2$\pi$ in the rest frame it has also completed one full 
revolution cycle from a laboratory point of view; so the spin-tune is a frame invariant quantity. 

For $p,p$ scattering we are not concerned with full turns in a magnetic storage ring; we are concerned 
with ``partial turns'' of one proton in the electric field of another---with the further complication 
that the ``other'' proton recoils. Roughly speaking though, in deflecting through an angular arc of, 
say, $2\pi/6$ the spin of a low energy particle in an electric field precesses relative to its momentum 
by $-\pi/3 = -60^{\circ}$.    

$G$-values for most of the stable atomic nuclei in a few upper rows of the periodic table, plus tritium, which is 
weakly unstable, are given in Appendix~\ref{sec:MDMs}, along with other parameters.  As mentioned previously,
$g$ and $G$ are approximate rational ratios for convenience for their use in precision determination
based on frequency domain determinations.
\footnote{For reasons that may or may not be physically significant, these $G$-values are also expressed as rational
fractions, with the integer ``6'' favored, meaning that the denominator is either less than 7, or a multiple 
of 6. In some cases the rational values are smaller than the measurement errors; in most cases the
rational fraction approximations are accurate to $\pm$1\%---good enough for semi-quantitative purposes,
such as estimating the number of turns for a spin to return to its initial orientations.  The number
``6'' is favored because the most favorable planar lattice is hexagonal for many purposes.  
Though this may be only of mystical significance, the rational fractions produce convenient answers.}

\subsection{T-violation and identical particle treatment}
In a magnetic storage ring (possibly with weak electric bending superimposed) independently polarized bunches 
collide at the crossing point.  Both initial proton polarization states are pure and 
some of the final state proton polarizations are measured with analyzing power close to 100\%.  
Quantum mechanical predictions will be tested by detecting correlation between the $p$-carbon measured spin 
orientations of final state protons.

Investigation of the influence of T-violation on $p,p$ elastic scattering is emphasized in this paper.
The combination of superimposed electric and magnetic bending and the measurement of final 
state polarizations enables the direct measurement of P- or T-violating amplitudes, which might 
very well have canceled in all previous experimental investigations of T-violation in $p,p$ scattering.  

The presence or absence of T-violation in nuclear forces is thought to bear significantly 
on important cosmological issues, especially missing mass, dark energy, and the matter/anti-matter 
imbalance.  The possible existence of a semi-strong, T-violating, nuclear force with coupling strength 
comparable to the electromagnetic interaction was proposed independently by Lee and 
Wolfenstein\cite{LeeWolfenstein}, by Prentki and Veltman\cite{PrentkiVeltman}, and by Okun\cite{Okun.1}\cite{Okun} in 1965. 

Unlike fixed target experiments, rather than being collinear, in Derbenov\cite{Derbenev} geometry, incident beams 
collide at right angles.  Observed in the laboratory frame, all incident and elastic scattered energies are equal.  It will
be shown that this provides a huge statistical polarimetric advantage.  \emph{Persuasive visual evidence of T-violation 
will be provided by unexpected correlation between the polarimetric $p$-carbon scattering directions of final state protons.}

Spin dependence is most easily detectable at low proton energy. The proposed COSY lab rearrangement is a 
variant of PTR\cite{CYR}, a prototype EDM measurement ring.  Rearrangement of existing COSY components into 
a ``FIGURE-8'' storage ring allows diametrically opposite polarized proton bunches in a single stored beam to 
collide.  ``Derbenev spin transparency\cite{Filatov}'' in figure-8 geometry is used to enable Fourier enhancement of 
T-violation sensitivity.  

Of the uncertain properties of nuclear physics, none is more fundamental than nucleon, nucleon interaction,
especially below pion production threshold.  The Derbenov collider configuration enables measurement of 
``elastic'' $p,p$ scattering spin dependence not possible using fixed polarized hydrogen target. To be consistent 
with previous discussion, these events would be ``prompt''.

This paper describes an experiment to investigate possible violation of time reversal invariance (T-violation)
in ``elastic'' $p,p$ scattering, where the quotation marks acknowledge the possibility of collisions for 
which the energy dissipation is undetectably small.  This applies especially to below threshold (closed)
resonant nuclear channels which would result in inelastic $p,p$ scattering at higher energy but which, being 
below threshold energy, produce no new secondary particles.  Above nuclear transmutation threshold energy 
(but below pion production threshold) these amplitudes promise substantial resonant room temperature 
storage ring nuclear transmutation.  This is to be investigated in a separate paper.

Derbenev et al.\cite{Derbenev} have produced a configuration enabling the investigation of scattering  
of orthogonal traveling protons at the crossing point of a figure-8 storage ring.  A solitary stored 
beam ``collides with itself'' in the sense that diametrically-opposite bunches automatically pass through each other in synchronism
at the intersection point (IP) where the beam crosses itself.  

There is a significant formal (but potentially qualitative) distinction between Wolfenstein and Derbenev geometry,
closely correlated with the treatment of identical particles and with issues of distinguishing time priority and left/right 
scattering ambiguity.  This ambiguity may correspond to a CM ambiguity noted by Ashtekar, De Lorenzo, and 
Khera\cite{AshtekarDeLorenzoKhera}

In (collinear-incidence) Wolfenstein geometry, for exactly-parallel, but transversely separated by 
immeasurably small distance, it is impossible to associate final state left and right scatters with 
incident-state left/right orbit displacement.  This is the source, in quantum mechanics, for treatment
of identical particle scattering, to sum amplitudes over interchanged incident states, before calculating intensities.

In Derbenev geometry there is less ambiguity.  A particle scattering promptly  
to the right must have come from the beam incident on the right, and vice versa.  This is illustrated in
Figure~\ref{fig:EikonalTrajectories}, to be discussed in more detail below.  One notes, based on the ray tracing shown,
that of the four possible final state angular quadrants, a promptly scattered beam particle from the left, say west (W),
may appear in the SW, NW, or NE quadrant, but be ``captured'' into a compound nucleus, preventing its prompt
escape into the SE sector.  This evolution would presumably proceed too quickly for scattering directions to be
correlated with scattering times.  Furthermore the time evolution description must be consistent for both scattered particles.
Nevertheless, the evolution can be modeled theoretically, no matter how quickly it transpires, and the calculated
distributions compared with experiment.  It is only its spin state that can enable a particle to ``remember'' 
whether it came from the left or from the right, based on its theoretical evolution.

In the fullness of time such ``resonant'' states can be expected to produce more nearly isotropic, but stochastically 
distributed scattering directions, which classical mechanics cannot handle, even in principle, and quantum mechanics can 
handle only probabilistically.  
Since energy and momentum need to be conserved ``in the long run'' the time delay between prompt and delayed scatters
may be immeasurable.  One conjectures that this is controlled by entanglement of the final states and detectable only
by measuring the correlation of final state spin orientations.  Behavior during The delayed time interval is not 
essentially different from an electron being captured stochastically into eigenstates sorted by energy and angular momentum;
except here it is stochastic decay into measurable final state protons.

For spin-1/2 particles, one conjectures that this distinction is correlated with the fact that in-plane 
(i.e. horizontal) deflections) preserve up or down states spin but mix horizontally polarized states.  One
also conjectures that these issues are correlated with rest-frame orientation ambiguity, which
implicates the sub-group structure of inhomogeneous Lorentz groups.  Otherwise it is difficult to
reconcile incident (and outgoing) beam directions being orthogonal in the laboratory, 
yet collinear in the rest frame.

In Wolfenstein geometry this
ambiguity is especially acute because the incident orbits are collinear.  In Derbenev geometry 
incident orbits can be coplanar but not collinear.

Accepting the Derbenev accelerator physics analysis of orbit and spin evolution, including ``spin transparency'', 
a figure-8 storage ring having polarimetry with nearly full $4\pi$ acceptance, and 
analyzing power ranging from 80\% to 100\% over substantial solid angular range is described.  Spin transparency 
enables a Fourier statistical enhancement of symmetry violation in $p,p$ scattering.

\subsection{Modern spin control; ancient nuclear physics}
The energy region emphasized in this paper, ``above'' Rutherford scattering, ``below'' meson threshold,
is paradoxical in various ways.  Total $p,p$ cross sections are discussed and plotted in detail
in a heroic 1993 review by Lechanoine-LeLuc and F. Lehar\cite{LeLuc-Lehar}, containing seven pages of 
references, from an era in which a large experimental group had five members.

\begin{figure}[hbt]
\centering
\centerline{\includegraphics[scale=0.38]{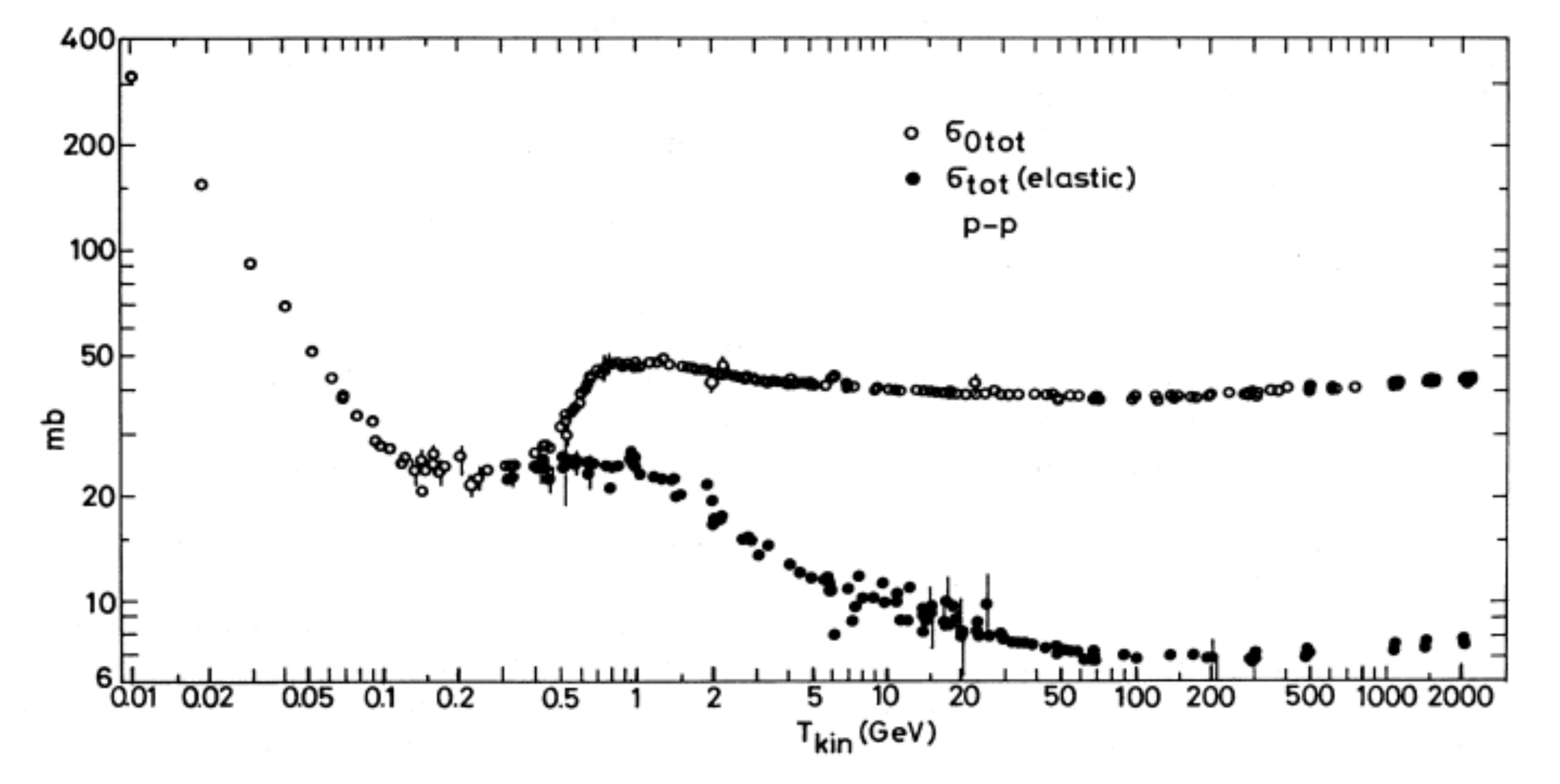}}
\centerline{\includegraphics[scale=0.40]{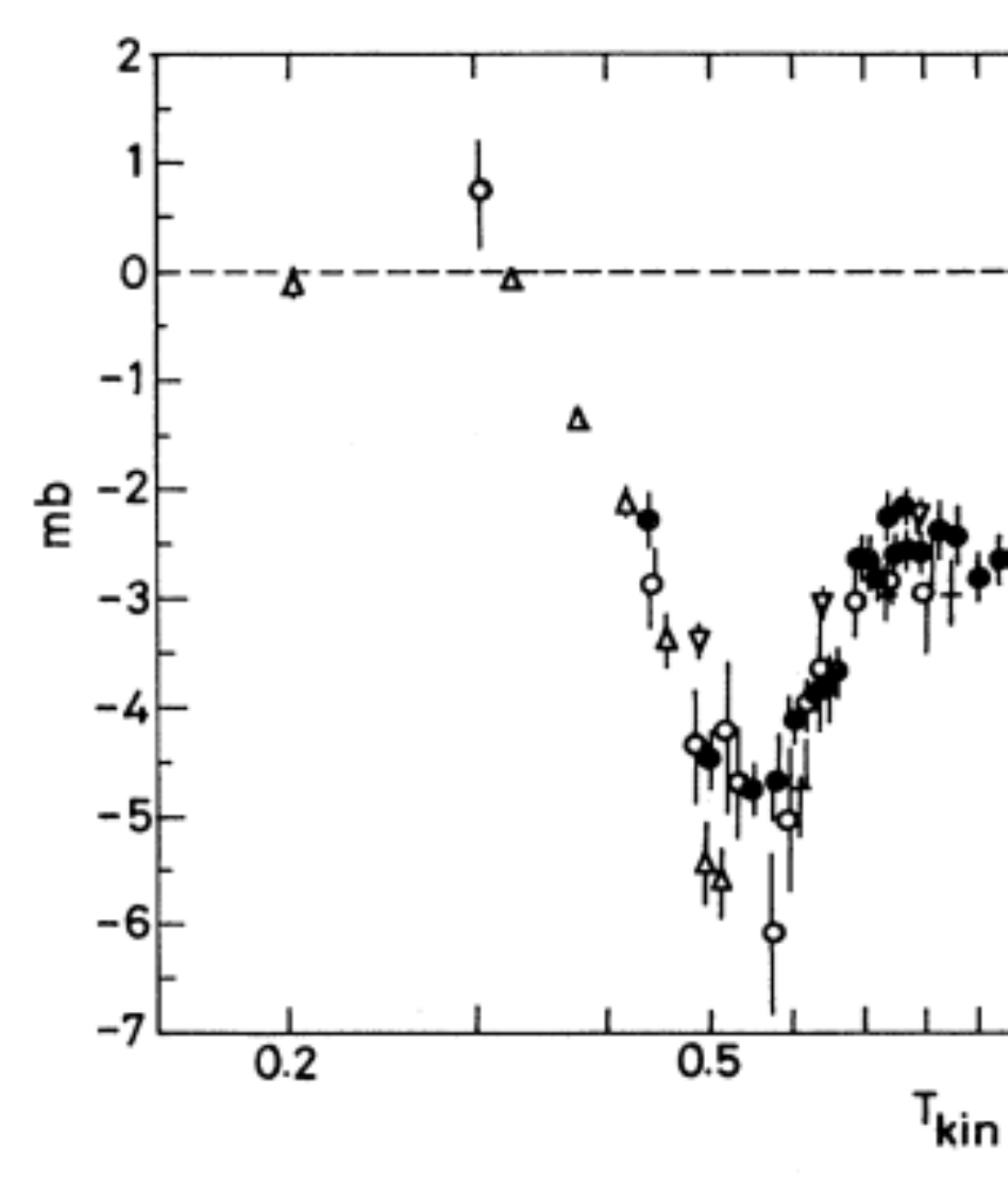} \includegraphics[scale=0.40]{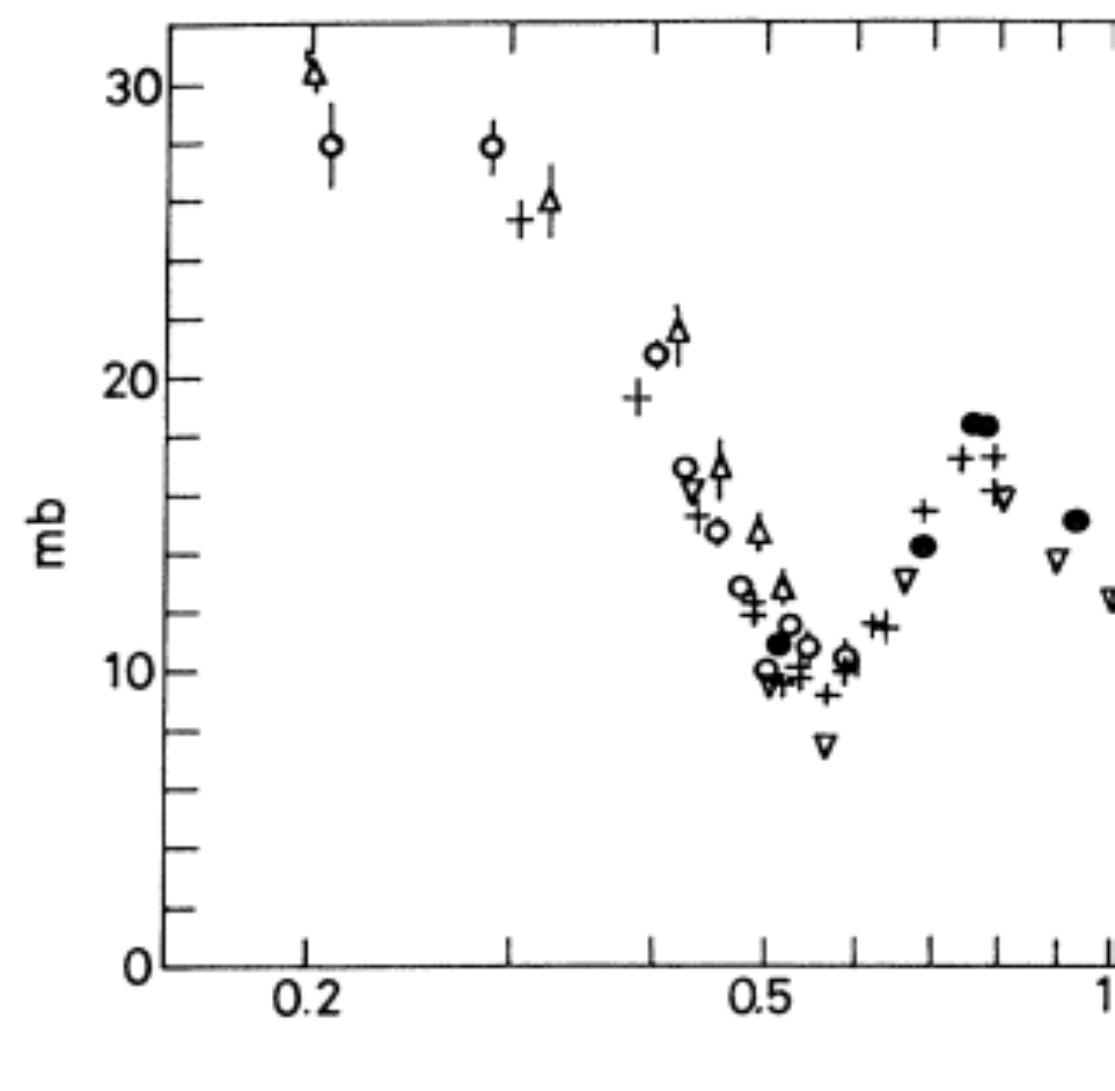}}
\caption{\label{fig:pp-total-crsctn-vs-KE}{(Copied from reference\ \cite{LeLuc-Lehar}) 
{\bf Above:\ }Energy dependence of $p,p$ spin-independent total cross section (open circles) and total 
elastic cross section (solid curves).}
{\bf Below left:\ }$\sigma_{1 tot}(p,p) = -(1/2)\Delta\sigma_T(p,p)$ energy dependence,
{\bf Below right:\ }$-\Delta\sigma_L(p,p)$ energy dependence on the incident state polarizations.
This figure should be compared with Figure~\ref{fig:deBroglie-lambda-vs-E_tot}.
}
\end{figure}

Figure~\ref{fig:pp-total-crsctn-vs-KE} (copied from LeLuc and Lehar) shows measured elastic and inelastic 
$p,p$ cross sections. Spin dependent $p,p$ cross sections, measured with polarized beams and polarized hydrogen 
target are plotted at the bottom of Figure~\ref{fig:pp-total-crsctn-vs-KE}. 

Though originally mysterious, the complicated long de Broglie wavelength behavior (i.e. low energy region below, 
say, 100\,MeV) quickly became well understood in terms of interference between Rutherford and nuclear 
amplitudes.  The same cannot be said for the short wavelength, higher energy region, above, 
say, 400\,MeV, where inelastic scattering quickly becomes dominant.  
\hspace*{-1cm} 

\subsection{Conjectured ``prompt'' $p,p$ scattering}
From an experimentalist's kinematic planning perspective $p,p$ elastic scattering cross section 
in the 100 to 400\,MeV energy range could scarcely be more boring---the scattering is isotropic and the 
total cross section is constant.  It seems curious, therefore, that the error bars and the scatter of 
values in this region are some five times greater than at almost any other region of the plot.  Though not 
at all suggested by the upper plot, an empirical, partial wave fit at the center of the region, at 310\,Mev, 
based on data available in 1965\cite{MottMassey}, required 12 non-vanishing partial waves, when the 
isotropy observed in coarse early experiments suggested that just one or two partial waves might 
be expected.  

My picture has the variation through this intermediate energy range representing the
transition from predominantly prompt, to primarily delayed, compound nucleus formation scattering 
probability, and ultimately to  pion production, when the nuclear forces cannot handle the energy. 

Of all nuclear processes it might initially have been expected that understanding elastic $p,p$ scattering 
would be assigned highest priority.  Based on counting citations, the number of PhD theses implied by the 
seven pages of reference\cite{LeLuc-Lehar}, this effort was strongly supported for several decades.  
Yet, it seems fair to say that $p,p$ elastic scattering is still not understood in any fundamental sense.

The plan is to improve this paradoxical history by better studying the influence of proton spins
by improving the previously feeble experimental capability to control initial spin states and the 
previous inability to measure final state polarizations.  

Since the angular momentum associated with the spins of fundamental particles is still mysterious, it 
should perhaps not be surprising that partial wave analysis, which is based on conservation of angular momentum
remains purely phenomenological.  The present proposal hopes to improve this situation.

Methods now available enable the preparation of pure initial spin 
states along with nearly ideal final state measurement of both final spin states over a significantly large
fraction of available phase space.  To be consistent with the huge investment in this 
area in the past one expects, therefore, significant support for these greatly improved methods in the present.

There is at least one exception to the statement made previously that $p,p$ scattering is not understood in any 
fundamental sense.\footnote{This GDH digression is the result of numerous conversations between the author and 
Kolya Nikoliev.}
It is the so-called Gerosimov-Drell-Hearn sum rule\cite{DrellHearn}\cite{Gerasimov},
which connects static properties of the nucleon---like the anomalous magnetic moment and the nucleon 
mass---with an integration from zero to infinity, of a difference of spin dependent doubly polarized total 
absorption cross sections of real photons.  Quoting from Helbing\cite{Helbing} ''The experimental data verify 
the GDH Sum Rule for the proton at the level of 8\% including the systematic uncertainties from extrapolations 
to unmeasured energy regions.''  Based purely on sound field theory, this surely qualifies as ``some kind of 
understanding in a fundamental sense''.
\footnote{As it happens, a peripheral capability of the apparatus we propose is the capability of precise measurement of MDMs
of the nuclei of many low mass isotopes, stable or weakly unstable, to which other sum rules might apply.  
Since the proton MDM is already known to 11 decimal places, it cannot be claimed that this MDM capability is required to
evaluate the right hand side of the GDH sum rule.  Neither may it be claimed to very much improve the accuracy of 
the integrand on the left hand of the GDF formula.}

What may be hoped for from the GDH sum rule is better understanding of the paradoxical $p,p$ interaction behavior under 
discussion.  From a qualitative perspective, one anticipates an improved theoretical understanding of how cross 
sections well outside our narrow energy range can help to understand seemingly elementary elastic $p,p$ scattering.

A paper by Bystricky, Lehar and Winternetz\ \cite{BystrickyLehar.1} provides a detailed 
breakdown of $p,p$ and $p,d$ scattering amplitudes, with special concentration
on time reversal invariance (TRI).
The paper by Lechanoine-LeLuc and Lehar\cite{LeLuc-Lehar}, whose importance has already been emphasized, 
reviews the voluminous nucleon-nucleon elastic scattering data, including polarization, with analyses, as of 1993, 
from well below to well above the energy range considered in the present paper, up to a few GeV.

Vastly more powerful experimental tools are available today than in the past. This is especially 
true of the precision spin control techniques that have been developed at the COSY laboratory in Juelich 
Germany\cite{Wilkin}\cite{Eversmann}\cite{Hempelmann}\cite{Rathmann-Kolya-Slim}\cite{Slim-Rathmann}\cite{RathmannWienFilter}.\\

The exchange force, postulated initially by Heisenberg, later revised by Wigner and others, 
models the dependence on the same spins that control the polarizations of elastic nucleon scattering. 
Along with the importance of exchange forces in fitting the binding energies of all nuclei from A=1 to 
A=200, it was the assumption of time reversal (T) conservation that constrained this modeling.  Then 
till now, other than its theoretical elegance, there has been no persuasive experimental evidence, 
one way or the other, for requiring T-conservation of the nuclear force.

With the anticipated EDM statistical precision inversely proportional to the square root of run duration
in mind, it has, until recently, been assumed that polarized beam run duration times would be limited by the 
spin coherence time (SCT).  Recent developments\cite{RT} have suggested that this will not be the case.  Run durations
in excess of 1000\,s, (a conventionally-adopted design goal) are now thought to be allowed
by lattice design optimized for the cancellation of sources of polarization decoherence.  

A more important, and unavoidable, consideration limiting run durations, is now thought to come from the 
consumption of beam particles associated with the destructive polarization measurements needed for phase 
locking the beam spin tunes\cite{RT}.  This capability is needed to enable electric and magnetic fields to 
be repeatably set, reliably reversed and reset, without the need for (unachievably precise) electric 
and magnetic field measurement.  

The ability to precisely reverse beam circulation directions provides powerful capability for reducing 
systematic error by averaging over beam revolution reversals.  This has lowered the priority associated 
with requiring two beams to counter-circulate simultaneously rather than consecutively.

These considerations have motivated, in the present paper, the re-ordering of PTR prototype ring development 
sequencing from the ordering defined initially in the CERN Yellow Report (CYR)\cite{CYR}.  Development
of the superimposed electric and magnetic sector bends needed for PTR can proceed ``in parallel''
with the construction of the FIGURE-8 ring and with the construction of the polarimetric tracking chamber 
needed for $p,p$ scattering measurement, and applicable as well to EDM measurement; all with the common 
goal of investigating time reversal symmetry.

\subsection{Investigation of the strong nuclear force}
In 1950, when 20\, MeV was ``high energy physics'', nuclear EDMs were thought to play a significant
role in the photo-disintegration of nuclei such as the deuteron.  This physics is explained in
Section XII-E of Morse and Feshbach\cite{Morse-Feshbach}.  The inferred value of the deuteron EDM 
at that time, as quoted by Gamow's Table I\cite{Gamow}, expressed as a photo-dissociation cross section was 
$2.7yimes10^{-27}\,{\rm cm^2}$.   The presence of what seems to be violation of time reversal (T) 
symmetry seems to have been forgotten.  Perhaps Norman Ramsey's surprise concerning a non-zero EDM effect 
provided some of his motivation for initiating his program to measure the neutron EDM? But subsequent 
measurements have provided only upper limits. 

With co-author Kolya Nikolaev, a Snowmass 2021 presentation\cite{Talman-Nikolaev} describes how a
ring such as described in the present paper can be used to investigate the possible existence of a
beyond standard model (BSM) \emph{semi-strong $T$-violation in elastic $p,p$ or $p,d$ scattering.} 

In $\tau$-$\theta$ puzzle days (late 50's to mid-60's) such a mechanism was suggested independently 
by Lee \& Wolfenstein\cite{LeeWolfenstein}, by Prentki \& Veltman\cite{PrentkiVeltman}, 
and by Okun\cite{Okun.1}\cite{Okun}\cite{KolyaFrankPaulo}.  Other, similarly motivated, conjectures have been numerous. 
Search for such a medium-strength, T-violating nuclear force provided the initial motivation for this paper.

The following few paragraphs have been extracted almost verbatim from Mott and Massey\cite{MottMassey}

\begin{figure}[hbt]
\centering
\includegraphics[scale=0.2]{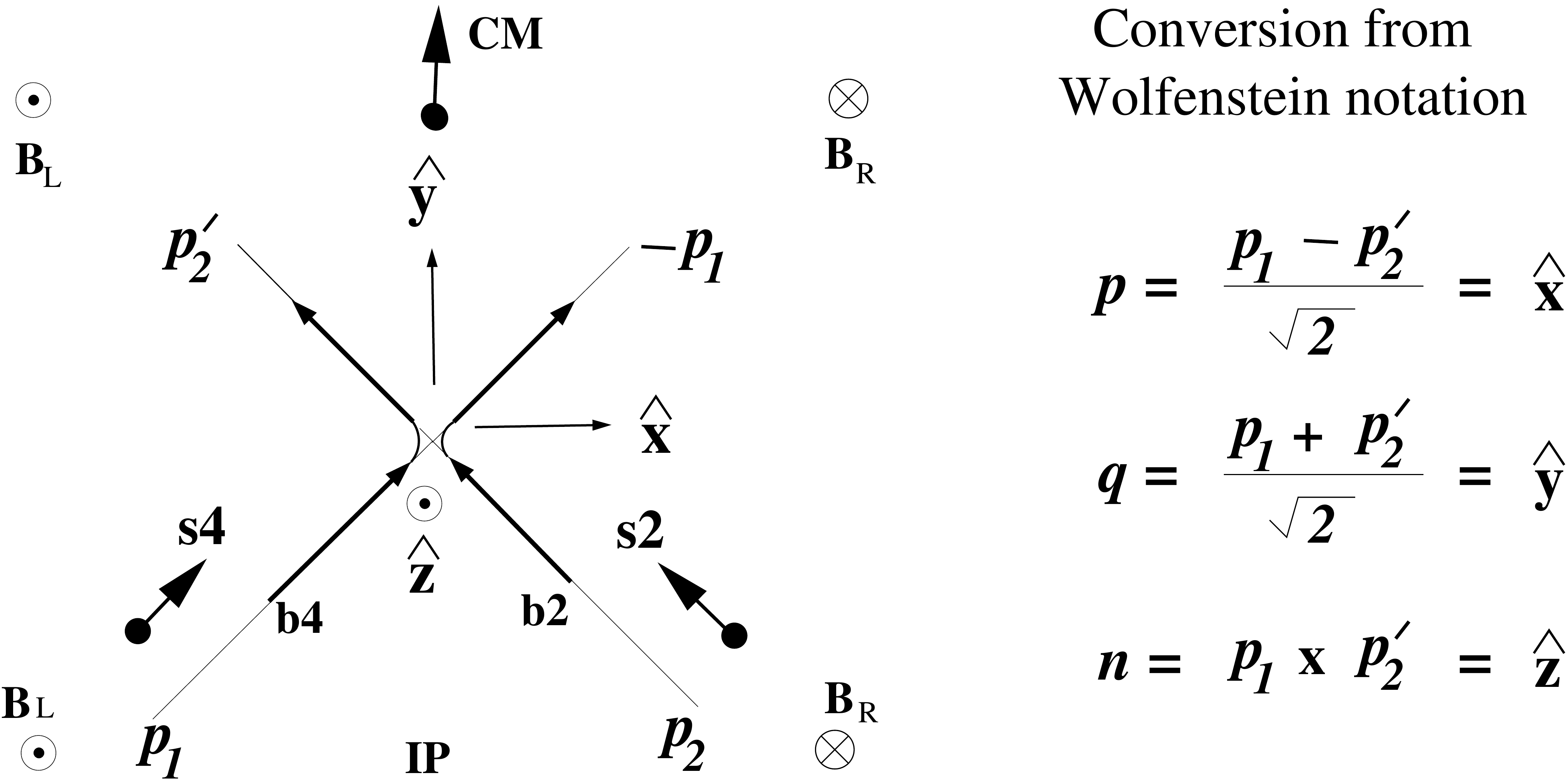}
\caption{\label{fig:WolfensteinAxes}For orthogonal beam collisions
the Wolfenstein incident momentum unit vectors ${\bf p}$ and ${\bf q}$, along with
normal to the scattering plane ${\bf n}={\bf p}\times${\bf q}, become
Cartesian unit vectors ${\bf\hat x}$, ${\bf\hat y}$, ${\bf\hat z}$,
in the transition from Wolfenstein to Derbenev configurations.
The ${\bf B_L}$ and ${\bf{ B_R}}$ symbols can only be understood in connection with 
Fig.~\ref{fig:Equal-period-pp-collider-mod-mod}.
}
\end{figure}

Especially influential were the ``scalar'' Wolfenstein operators, for primary beam particle ``1'' 
secondary target particle ``2''; 
$$
\pmb{1},\ \
\pmb\sigma_1\cdot\pmb\sigma_2, \ \ 
(\pmb\sigma_1+\pmb\sigma_2).{\pmb n}, \ \ 
(\pmb\sigma_1-\pmb\sigma_2).{\pmb n}, \ \ 
$$
\vskip -1cm
$$
(\pmb\sigma_1\cdot{\pmb p})(\pmb\sigma_2\cdot\bf p),  \ \ 
(\pmb\sigma_1\cdot{\pmb n})(\pmb\sigma_2\cdot\bf n),  \ \ 
(\pmb\sigma_1\cdot{\pmb q})(\pmb\sigma_2\cdot\bf q) \ \ 
$$
where ``\pmb{1}'' is the identity matrix, and
the three components of the ${\pmb\sigma_1}$ and ${\pmb\sigma_2}$ ``vectors'' are the three Pauli 2x2 matrices.
As shown in Fig.~\ref{fig:WolfensteinAxes},
orthonormal coordinate basis vector ${\bf n}$ is an axial unit pseudo-vector which, with
orthogonal incident momentum unit vectors ${\bf p}$ and  ${\bf q}$, defines the scattering plane, to which ${\bf n}$ 
is orthogonal.  These are the possible forms in terms of which the scattering matrix can be composed and matched
phenomenologically with measured values at each value of beam energy. 

Already formidably complicated, two Wolfenstein ``pseudo-scalar'' forms,
\begin{align}
 (\pmb\sigma_1&\times\pmb\sigma_2)\cdot{\pmb n},\ \hbox{and} \label{eq:Violate.1}\\ 
(\pmb\sigma_1\cdot{\pmb p})&(\pmb\sigma_2\cdot{\pmb q}) + (\pmb\sigma_1\cdot{\pmb q})(\pmb\sigma_2\cdot{\pmb p})
\label{eq:Violate.2}
\end{align}
have (conventionally) been excluded on the basis that they violate T-reversal and/or P-symmetry.
An intermediate strength T- or P-violating nuclear force can be represented by a superposition of
operators~\ref{eq:Violate.1} and  \ref{eq:Violate.2}.

Based on this phenomenological formulation, a description of nuclear forces capable of accurately accounting 
for the parameters of all nuclei and of accounting for the decay of all unstable nuclear isotopes
was gradually established. A substantial portion of this understanding was based on the treatment of protons 
as elementary particles for which there was no relevant concept of ``binding energy'' required.  Binding 
energies were defined relative to the proton binding energy (taken to to be zero).  In detail, the exchange 
force was taken to be a best compromise of Heisenberg, Wigner and Majorana exchange force variants, based 
on measured binding energies. 

By $\tau$-$\theta$ puzzle days,  the elementary particle community had ``moved on'' from what had by then become 
``low energy'' nuclear physics, accepting, with little subsequent alteration, the existing phenomenologic description 
of the force between nucleons.  Note, however, that there has been little other than theoretical prejudice for 
requiring time reversal symmetry conservation.  

In the meantime, Gell-Mann quarks had been introduced as constituents of the proton.  This made protons 
composite, contrary to already adopted modeling that treats the proton as elementary.  
Logically, its composite nature  might reasonably have been accompanied by the possibility of the proton itself 
having binding energy\footnote{In a paper introducing integer charge quarks, Han and Nambu\cite{HanNambu} suggested 
implicitly the influence on binding energies, but, with ``color'' persuasively introduced, this seems not to have 
been pursued.}.

There is thought to be a close connection between time reversal (T-symmetry) and the baryon/anti-baryon 
imbalance in our
present day universe.  This connection was stressed by Sakharov\cite{Sakharov} in an early and influential
paper.  Nowadays, to a first approximation, our universe contains only protons.  One has to suppose that the laws 
of particle physics are consistent with symmetry-breaking processes which have resulted in a surplus of what we 
now call protons.  As Sakharov famously explained, any generation of such an imbalance of particles 
and anti-particles required statistical disequilibrium conditions, absence of baryon number conservation,
and CP- (and hence also T-) violation in one or more of the fundamental laws of particle production and decay.  

Building on the Wolfenstein formulation, traditional analyses, after having dropped the T-violating operators,
have proceeded to simplify the algebra by proceeding to a density matrix formulation.  This approach 
assumes implicitly that every experimental elastic scattering  apparatus is limited in a way that,
for identical particles, requires averaging over incident states and/or summing over final state amplitudes 
with interchanged output directions.  Here it is argued that the summing of amplitudes with interchanged 
output directions is not appropriate in Derbenev geometry, since, for the majority of them, it is possible to 
match each promptly scattered proton with the incident beam, left or right, that it came from.

The fact that the Pauli operators are 2x2 matrices makes it difficult to acquire any simple intuitive 
correlation between  momenta, on the one hand, and separation into spin-flip and non-spin-flip events on the other.
Density matrix formulation makes this algebraically convenient.  This is even true when both incident states are pure,
as we are assuming.  Mott and Massey\cite{MottMassey}, Section~IX.5, explain this in detail.  It is useful
to employ spherical coordinates with in-plane cylindrical polar angle $\theta$ and out of plane azimuthal $\phi$ 
coordinates. 

For the detection apparatus proposed here there is no need for the summing or averaging of amplitudes.  
The explicit isolation of symmetry-violating and symmetry-conserving amplitudes makes the basic Wolfenstein 
formulation seem appropriate.  However, the Derbenev geometry also avoids the need for averaging over final 
states to account for identical incident particle types.  Clearly, in an actual elastic nuclear scatter, 
each strongly scattered proton counts in the sector at which it \emph{was not previously aimed}.  

The Wolfenstein configuration suffers from the essential singularity appearing at the center of phase space which 
makes particle interchanged summation necessary.  In the Derbenev configuration this identical particle
singularity appears harmlessly in the form of weakly scattered proton directions centered on the forward direction 
but with large impact parameters approaching the boundary of phase space.

\subsection{Importance of anomalous nuclear MDM $G$-values\label{sec:AnomMDMs}}
The motivation for the storage ring or (or rings) being promoted in this paper centers on the careful study of
``elastic'' nucleon scattering.  The guiding rationale emphasizes the important role played by the  anomalous 
MDM, $G$, in elastic nucleon scattering.  An essential feature of the rings being advocated here 
follows from  their superimposed electric and magnetic bending, which provides the capability of 
simultaneously counter-circulating frozen spin beams of different particle type.  

As explained in 
publications\ \cite{RT-ICFA} and\ \cite{RT-CLIP-PTR}, this storage ring configuration is ideal for 
the precision measurement of anomalous nuclear MDM $G$-values. Such rings serve naturally for the function of 
``mutual co-magnetometry'' for precise measurement of the ratios of the numerator and denominator integers entering 
into the precision experimental determination of $G$-values of the particle types in the two counter-circulating 
beams.  

For historical reasons, based probably on the great importance and successful application of the $g$-factor in 
atomic physics, the (dimensionless) parameter $G$ is considered to be subsidiary to the (also dimensionless) 
``$g$-factor'', which expresses (in our nuclear case) a fundamental measurable ratio of nucleus angular momentum 
(proportional to inertial mass $m$ of nucleon) to magnetic moment (proportional to charge of the same nucleus). 

With $Z$ and $A$ being dimensionless measures, the ratio of integers, $A/Z$, justifies regarding   
$g(A/Z)$ as function $A$ and $Z$ only via the ratio $A/Z$.  To be ``anomalous'' the 
dimensionality of $G$ and $g$ must be the same: i.e. dimensionless.  

For every nucleon, $Z$ is truly an integer 
multiple of (positive) proton charge $e$.  Regrettably, for example because of nuclear binding energy, nucleon 
mass ratio is only only approximately given by the mass number $A$.

This discussion is continued in Appendix~\ref{sec:MDMs}, but not because it is of lesser importance;
in fact this paper provides further strong support for the precise measurement, and consistency
treatment of nuclear isotope MDMs and mass values. But the discussion is both technical and boring.  
This justifies treating Appendix~\ref{sec:MDMs} as a self-contained discussion of the experimental and 
theoretical connections between $g$ and $G$\cite{NIST-background}\cite{NIST-isotope-abundance}.

\begin{figure}[hbt]
\centering
\includegraphics[scale=0.9]{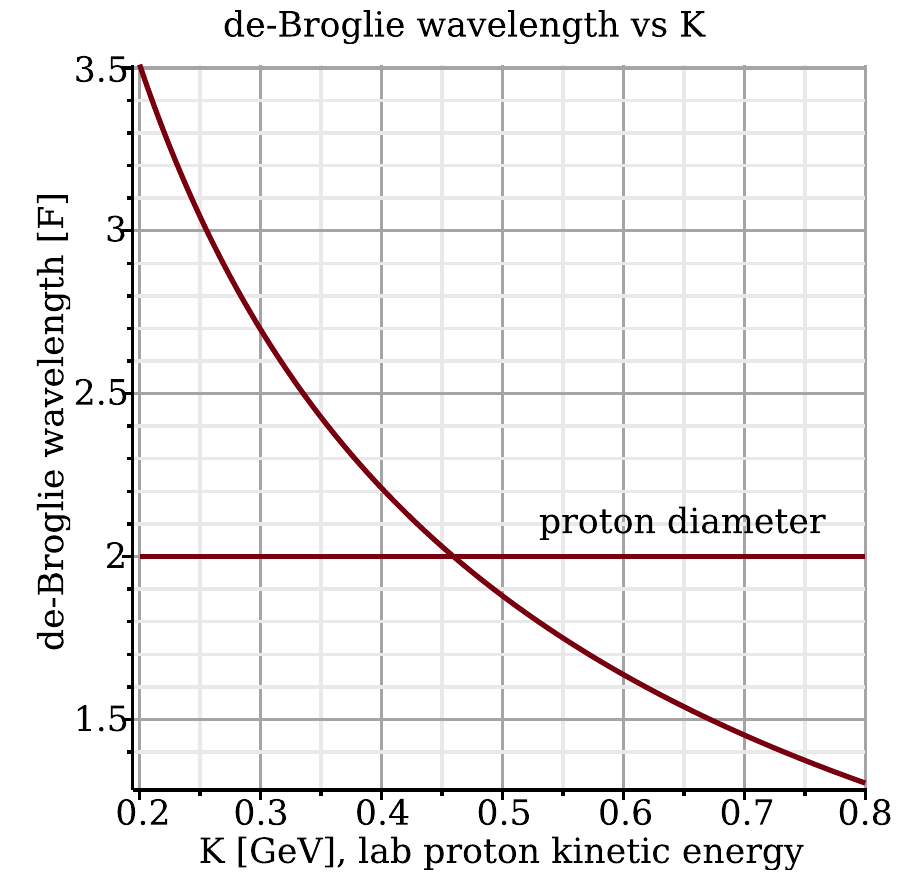}
\caption{\label{fig:deBroglie-lambda-vs-E_tot}Fermi length units plot of the deBroglie wavelength of an 
incident proton versus its total laboratory kinetic energy.  See Appendix Section~\ref{sec:Orders-of-magnitude}, 
for numerical values of various nuclear parameters. The approximate match of deBroglie wavelength with proton 
size can scarcely be simple coincidence. Surely this represents a nominal transition point from wave to
particle description of $p,p$ scattering; much like the transition from geometric to wave optics.
New ``particles'' need to be massive, in this case pi-mesons, to account for the available energy, while
conserving momentum at the local velocity.  Azimuthal symmetry is broken on an event by event basis, 
but is, presumably, preserved on average.  For correlation with, say, rainbows, 
refer to Van-de-Hulst\cite{Van-de-Hulst};  a further empirical numerical factor, $\pi(m-1)$ of order 1, 
multiplying the diameter, is required, where $m$ is ``index of refraction''.} 
\end{figure}

\section{Orthogonal collisions of paired bunches in a single beam figure-eight storage ring}
\subsection{Wolfenstein/Derbenev amplitude comparison\label{sec:CoordinateRotation-0}}
During the infancy of nuclear physics all possible experiments were ``fixed target'', with kinematically 
unbalanced initial states;  with special relativity still controversial, ``energy'' meant kinetic energy (K),
which, by convention, was used to label energy dependent data.  A century later, using Derbenev figure-eight 
geometry, we are in a position to perform the same definitive experiments with symmetric incident states.

Before continuing, it has to be recognized that what has been been referred to as WOLFENSTEIN configuration
has been specific in meaning ``collinear incident beams'', but ambiguous as to frames of reference.  Before
the advent of storage rings the incident beam passed through a target fixed in the ``laboratory frame'',
where unprimed kinematic parameters are employed.  In the center of mass system, whose parameters are 
distinguished from corresponding laboratory frame by the attachment of a single prime.  Though the
centroid moves in the laboratory, it stays on the original beam line and the centroid of every scatter 
starts on, and remains on, the same stationary beam line.  

Nowadays, for high energy circular collinear (Wolfenstein) colliding beam storage rings, mechanical energy 
$\mathcal{E}=mc^2+K$ is used to specify incident beam energies. This labeling is ideal, since the center of 
mass (CM) and laboratory frames are identical.  But, for Derbenev beam self-collision the CM and laboratory 
frames are different.  (The same is true for collisions between countercirclating in rings with
superimposed electric and magnetic bending.)  For meaningful comparison of D\&W physics results it is useful 
to introduce a ``comparison energy'' $M^*=\sqrt{s},$ which, in every case, is the total mass of the $p,p$ 
system in the CM reference frame.  The natural data comparison then is between data sets for which 
$$M^* =\sqrt{s} = M^*_W = M^*_D.$$ 

Laboratory frame descriptions are convenient for both W and D configurations.
This makes it sensible to perform all kinematic calculations in the laboratory before equating 
their determinations of $M^*$ which (with $c=1$) is the rest energy in the CM frame. All other
kinematic variables refer implicitly to the lab frame.  Avoiding kinetic energy (until the end) 
and never evaluating square roots, avoids the need ever to solve a quadratic equation. 

With ``W'' subscripts, the square of the total energy in collinear Wolfenstein fixed target
laboratory frame configuration, with $c=1$ and total energy squared $\mathcal{E}^2_{\rm tot.-W}$,
and with proton mass $m$, is given by
\begin{equation}
\mathcal{E}^2_{\rm tot.W} = (\mathcal{E_{\rm W}} + m)^2 
                         = \mathcal{E}_{\rm W}^2 + 2m \mathcal{E}_{\rm W} + m^2.
\label{eq:DW-1}
\end{equation}
Total momentum and CM rest mass are then given by 
\begin{align}
p^2_{\rm tot.W}  &= p^2_{\rm_W}, \hbox{\quad and} \notag\\
 {M^*_{\rm W}}^2 &= \mathcal{E}^2_{\rm tot.W} - p^2_{\rm tot.W} \notag\\
                &= \mathcal{E}_{\rm W}^2 + 2m \mathcal{E}_{\rm W} + m^2 - p^2_W \notag\\
                &= 2m \mathcal{E}_{\rm W} + 2m^3.
\label{eq:DW-2}
\end{align}
In Derbenev figure-8 configuration,
\begin{align}
\mathcal{E}^2_{\rm tot.D} &= (2\mathcal{E}_{\rm D})^2 = 4\mathcal{E}_{\rm D}^2,\notag\\
p^2_{\rm tot.D}           &= \Big(2\frac{p_{\rm D}}{\sqrt2}\Big)^2 = 2 p_{\rm D}^2 ,\notag\\
{M^*_{\rm D}}^2           &= \mathcal{E}^2_{\rm tot.D} - p^2_{\rm tot.D} \notag\\
                         &= 4\mathcal{E}^2_{\rm D} - 2 p_{\rm D}^2 \notag\\
                         &= 2\mathcal{E}^2_{\rm D} + 2 m^2. 
\label{eq:DW-3}
\end{align}
Equating centroid rest energies produces
\begin{align}
\mathcal{E}^2_{\rm D} = m \mathcal{E}_{\rm W}
\label{eq:DW=4}
\end{align}
The purpose for this calculation has been to determine, in Derbenev configuration, the ``ring energy'',
$\mathcal{E}_{\rm D}$, that corresponds to the Wolfenstein  configuration  ``laboratory energy'', $K_{\rm W}$,
in terms of which historical experimental data is still recorded and referenced.  
(This resembles the historical nuisance task needed to transform angular distributions measured using targets 
fixed in the laboratory, to center of mass angles.) 
 
Finally, to complete this task, requires relating $K_{\rm W}$, the ``historical kinetic energy''\footnote{Historically, $K_{\rm W}$ 
was expressed simply as ``particle energy''.} to ``modern day mechanical energy'', $\mathcal{E}_{\rm D}$;
\begin{equation}
 \mathcal{E}^2_{\rm D} = m (m + K_{\rm W}).
\label{eq:DW-5}
\end{equation}
This relation is troubling from both mathematical and quantum mechanical perspectives. 
With $m$ treated as necessarily real, the limiting behavior of 
${E}_{\rm D}$ and $K_W$, individually treated as analytic functions of $m$, the square root required in the 
transformation becomes mathematically problematic as $m\rightarrow 0$. 

From a quantum wave mechanical perspective, the issue is more confusing, especially with respect to 
deBroglie relations for momentum and energy, $$p=h/\lambda,\quad E=h\nu,$$ with all quantities in these
relations required to be real.
\footnote{The appearance of the symbol ``i'' as an abbreviation for $\sqrt{-1}$
enters QM formulas both for \emph{trivial convenience} in the expression of trigonometric identities, and for 
\emph{deep physics reasons} such as expressing the resonant build-up or exponential-decay of physical states.  
Strictly speaking, it is not logically required for just one symbol to play both of these roles.  However
this suspect symbolic usage is hopelessly entangled (pun intended) in customary treatments and understanding 
of quantum mechanics, including the present paper.}


For the 183.1\,MeV central energy of the Derbenev ring proposed in the present 
paper, the kinetic energy is obtained as;
\begin{align}
   & K_{\rm W} = (m + K_{\rm D})^2/m - m, \notag\\
   & \overset{e.g.}{\ =\ }\frac{(0.938 + 0.183)^2}{0.938} -0.938 = 0.402\,{\rm GeV},
\label{eq:NR-intro}
\end{align}
which is appropriate for the horizontal axis scales
in Figure~\ref{fig:pp-total-crsctn-vs-KE}.  Since the energies are only weakly 
relativistic, the Derbenev kinetic energies are roughly half the Wolfenstein kinetic energy.

\begin{figure*}[hbt]
\centering
\includegraphics[scale=0.35]{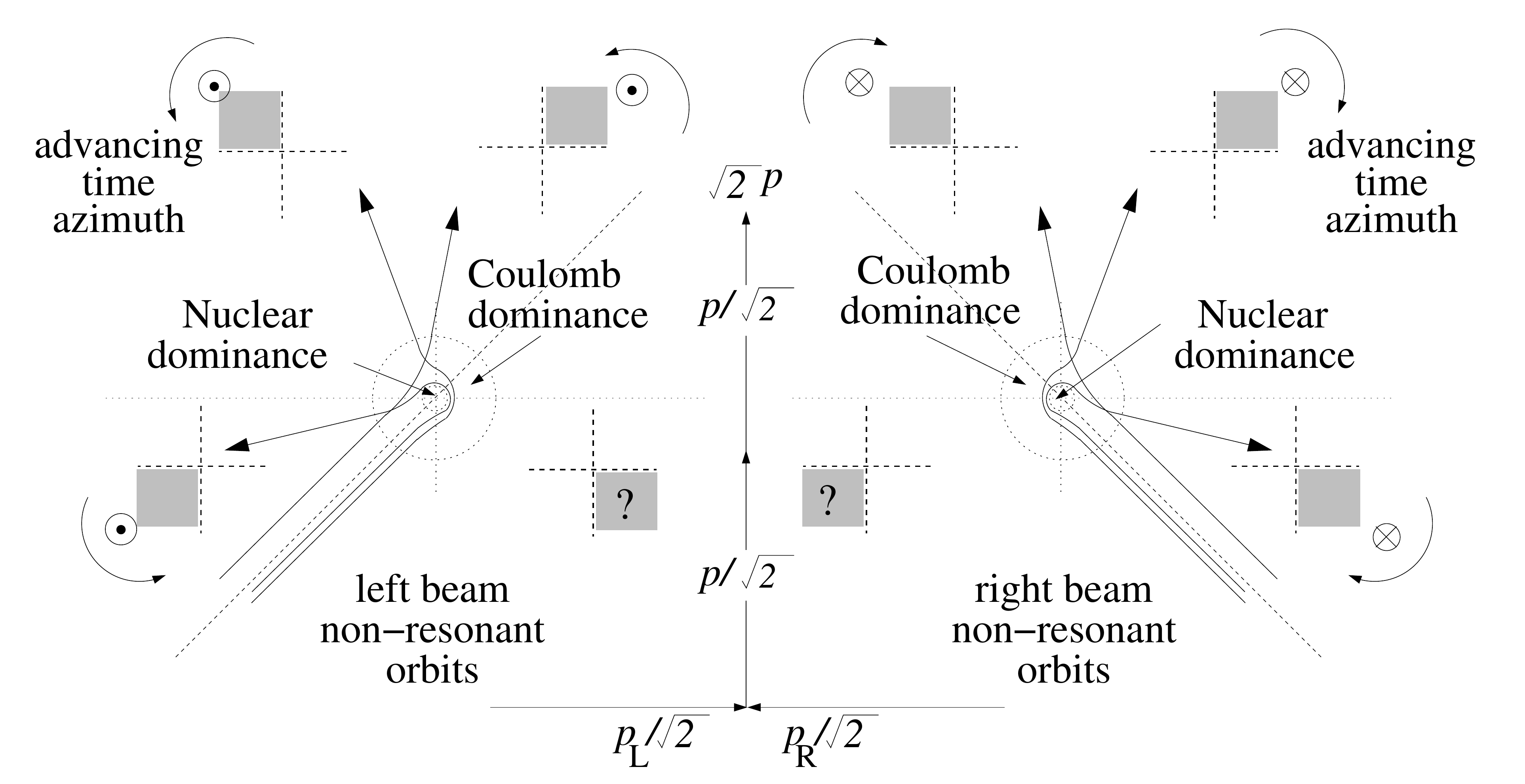}
\caption{\label{fig:EikonalTrajectories}''Prompt'' classical orbits (or wave eikonal curves) representing 
a sequence of three  particles (or wave packets) sequentially following the mutual scattering of a pair of protons, 
one from each beam, one into each of three azimthally (i.e. ``in a periodic angular sense'') advancing sectors. 
By the fourth sector it has become ambiguous (classically) or probabilistic (quantum mechanically) whether the particle 
is detectable immediately or has been  ``mutually captured'' by the 
other nucleus.  Semicircular arrows indicate the sense of azimuthal advance (which is necessarily opposite opposite for
the two scattering particles. 
The epoch represented begins just before and ends just after the actual collision.  Inset
graphs show horizontal plane quadrants, viewed (for example) from above. The shadings represent the sequencing of events, 
by indicating the quadrants in which successive events terminate.  Individual orbits orbits can be labeled by ``azimuthal time'',
with one ``typical'' prompt orbit shown in each of the first three quadrants after the collision. 
As regards Newtonian particle description this represents the ``complete story''---the subsequent evolution
is unpredictable in principle.  As regards ``quantum wave physics'' the figure represents evolution only 
up to the end of an initial time interval---the subsequent evolution is predictable in principal, but
only in a statistical sense---for example, the fourth scatter may, or may not, result in an event in the
fourth quadrant.  Calculation of the subsequent probabilities requires quantum mechanics, and is limited by 
the Heisenberg (time) uncertainty; or (equivalently) by limitations imposed by mathematical properties
of the Fourier transforms used to calculate the probabilities.}
\end{figure*}
%

The proposed figure-8 storage ring is shown schematically in Figures~\ref{fig:Equal-period-pp-collider-mod-mod}
and \ref{fig:CPT-FigureEight-layout}.  Though intended to serve as a measurement of elastic scattering, the experiment 
commences by treating the apparatus as two matched polarimeters being ``calibrated'' in coincidence. A single
beam contains four bunches, 
${\bf b_1}$, ${\bf b_2}$, ${\bf b_3}$, and ${\bf b_4}$, with individually controlled polarizations 
${\bf s_1}$, ${\bf s_2}$, ${\bf s_3}$, and ${\bf s_4}$.  Fig.~\ref{fig:Equal-period-pp-collider-mod-mod}
shows bunches ${\bf b_2}$, and ${\bf b_4}$ about to collide at the intersection point (IP).

It is implicit in the Wolfenstein operator definitions for the incident beams to be collinear.  
In our application the incident beams are orthogonal in the laboratory reference frame, even
though they are collinear in the CM frame, (which is shown moving upward in Fig.~\ref{fig:WolfensteinAxes}).  
Our configuration, which resembles Wolfenstein's $t$-channel, is shown with time advancing
directly up the page. 

The CM velocity is not very large in the laboratory, and significant CM symmetry is preserved 
because of the transversely symmetric orthogonal approach of the incident beams in the laboratory.  Though the
protons approach at right angles in the lab, in the CM frame the approach is collinear, as are the scattered 
trajectories.

The corresponding laboratory frame constraint is that incident and scattered momenta pairs are 
orthogonal, and all four energies are identical.  In the laboratory, as well as defining a (horizontal) plane,
the incident trajectories define a $\pm\pi/4$ ``right-angle cone'', centered on the transverse axis along 
which the CM is traveling.  The scattered trajectories necessarily lie on the other branch of the same cone,
necessarily orthogonal also, and defining a plane containing the same conical axis but, in general,
skew to the horizontal plane, by some non-zero azimuthal angle $\Phi$.  

With the incident particle momenta defining a circle on the cone of incidence, the scattered particle momenta lie
on the mirror-symmetric circle on the other branch of the cone.  As a consequence, the scattered particle trajectories 
are also orthogonal in the laboratory with identical energies, the same as the incident particle energies.  
\emph{It is their equal energies that permit both scattered particle polarizations to be measured with nearly 
perfect 100\% analyzing powers.}  

Wolfenstein's incident particle momenta are ${\bf p_1}$ and ${\bf p_2}$, and scattered particle momenta are 
${\bf p'_1}$ and ${\bf p'_2}$.  Our incident particles ${\bf p_1}$ and ${\bf p_2}$ are  
his ${\bf p1}$ and ${\bf -p'_2}$.  Fig.~\ref{fig:WolfensteinAxes} includes translation formulas on the right
for conversion back to Wolfenstein's basis vector definitions.

So far it is only the ${\bf b_2}, {\bf b_4}$ bunch pairing that has been discussed.  Of course, the same discussion
applies to the ${\bf b_1}, {\bf b_3}$ bunch pairing.  During bunch polarization preparation each of the four bunch
polarizations can be prepared arbitrarily.  This makes it convenient to compare the different outcomes from
differently prepared incident states merely by maintaining the separation of ``even bunch index'' and  
 ``odd bunch index'' incident bunch pairs.

\section{Paired bunch collisions in electric/magnetic storage rings}
\subsection{Superimposed magnetic/electric bending\label{sec:CoordinateRotation-2}}
During the infancy of nuclear physics all possible experiments were ``fixed target'', with kinematically 
unbalanced initial states;  with special relativity still controversial, ``energy'' meant kinetic energy (K),
which, by convention, was used to label energy dependent data.  A century later, using superimposed electric
and magnetic bending, we are in a position to perform the same definitive experiments with symmetric incident states.

Nowadays, for high energy circular colliding beam storage rings, mechanical energy $\mathcal{E}=mc^2+K$ is used for
incident beam energies. This labeling is ideal, since the center of mass (CM) and laboratory frames are identical.  
But, for scattering in the CM, but particle detection in the laboratory the frames are different.  
For meaningful kinematic comparison it is useful to introduce as ``comparison energy'' $M^*=\sqrt{s}$, which is 
the total mass of the $p,p$ system in the CM reference frame.

The proposed figure-8 storage ring is shown schematically in Figures~\ref{fig:Equal-period-pp-collider-mod-mod}
and \ref{fig:CPT-FigureEight-layout}.  Though intended to serve as a measurement of elastic scattering, the experiment 
commences by treating the apparatus as two matched polarimeters being ``calibrated'' in coincidence. A single
beam contains four bunches, 
${\bf b_1}$, ${\bf b_2}$, ${\bf b_3}$, and ${\bf b_4}$, with individually controlled polarizations 
${\bf s_1}$, ${\bf s_2}$, ${\bf s_3}$, and ${\bf s_4}$.  Fig.~\ref{fig:Equal-period-pp-collider-mod-mod}
shows bunches ${\bf b_2}$, and ${\bf b_4}$ about to collide at the intersection point (IP).

It is implicit in the Wolfenstein operator definitions for the incident beams to be collinear.  
In our application the incident beams are orthogonal in the laboratory reference frame, even
though they are collinear in the CM frame, (which is shown moving upward in Fig.~\ref{fig:WolfensteinAxes}).  
Our configuration, which resembles Wolfenstein's $t$-channel, is shown with time advancing
directly up the page. 

The CM velocity is not very large in the laboratory, and significant CM symmetry is preserved 
because of the transversely symmetric orthogonal approach of the incident beams in the laboratory.  Though the
protons approach at right angles in the lab, in the CM frame the approach is collinear, as are the scattered 
trajectories.

The corresponding laboratory frame constraint is that incident and scattered momenta pairs are 
orthogonal, and all four energies are identical.  In the laboratory, as well as defining a (horizontal) plane,
the incident trajectories define a $\pm\pi/4$ ``right-angle cone'', centered on the transverse axis along 
which the CM is traveling.  The scattered trajectories necessarily lie on the other branch of the same cone,
necessarily orthogonal also, and defining a plane containing the same conical axis but, in general,
skew to the horizontal plane, by some non-zero azimuthal angle $\Phi$.  

With the incident particle momenta defining a circle on the cone of incidence, the scattered particle momenta lie
on the mirror-symmetric circle on the other branch of the cone.  As a consequence, the scattered particle trajectories 
are also orthogonal in the laboratory with identical energies, the same as the incident particle energies.  
\emph{It is their equal energies that permit both scattered particle polarizations to be measured with nearly 
perfect 100\% analyzing powers.}  

Wolfenstein's incident particle momenta are ${\bf p_1}$ and ${\bf p_2}$, and scattered particle momenta are 
${\bf p'_1}$ and ${\bf p'_2}$.  Our incident particles ${\bf p_1}$ and ${\bf p_2}$ are  
his ${\bf p1}$ and ${\bf -p'_2}$.  Fig.~\ref{fig:WolfensteinAxes} includes translation formulas on the right
for conversion back to Wolfenstein's basis vector definitions.

So far it is only the ${\bf b_2}, {\bf b_4}$ bunch pairing that has been discussed.  Of course, the same discussion
applies to the ${\bf b_1}, {\bf b_3}$ bunch pairing.  During bunch polarization preparation each of the four bunch
polarizations can be prepared arbitrarily.  This makes it convenient to compare the different outcomes from
differently prepared incident states merely by maintaining the separation of ``even bunch index'' and  
 ``odd bunch index'' incident bunch pairs.

\subsection{Near perfect analyzing power}
Nearly perfect analyzing power is less impressive (but no less essential) than it seems.
As a proton slows down, its analyzing power exceeds 0.99 only briefly.  However the analyzing power remains greater
than, say,  0.8 for an appreciable fraction of the proton's full range. Since the scattered proton energies are identical
in the laboratory, the beam energy does not need to be much greater than 183.1\,MeV, for all scattered energies to 
exceed the energy at which the graphite analyzing power exceeds 99\%.  Also graphite chamber thickness 
great enough to stop all scatters is easily achieved, irrespective of scattering angle. 
See Fig.~\ref{fig:p-C-analyzing-power} and Table~\ref{tbl:graphite-stopping}.

The polarization of every proton scattered at the IP, and scattered again in one of the near crystalline graphite foil 
(or construction grade graphene) plates of the detection chamber, will have been determined with unprecedented accuracy, 
irrespective of scattered proton direction.  This will provide high quality positive elastic signature for every elastic 
scatter.  The total stopping ranges of most protons, including weakly inelastic scatters, will be used to reject 
inelastic scatters.

With such cleanly matched pairs of elastic scatters, the comparison of time-forward and (effectively) 
time-reversed scatters can be performed with unprecedented accuracy on an event by event basis. For example, 
the equality of scattering probability $P$ and analyzing power $A$ (required to be equal by time reversal 
symmetry) can be checked for matched pairs of protons.  Furthermore, (with sufficient incident 
spin orientation control) by altering the incident pure polarization states, the reversed-time version of 
any observed forward-time scatter can be re-created.  In other words, truly forward-time and 
backward-time scatters can be compared.

\section{Low energy $p,p$ elastic scattering search for time-reversal violation\label{sec:ppscat}}
\subsection{Detection apparatus}
The goal of the proposed scattering experiment is to reduce, by a large factor, elastic scattering $p,p$
T-violating upper limits, currently roughly one percent. This would use highly polarized beams in the
figure-eight storage ring being described in the present paper for colliding beam elastic scattering measurement.
Even though some electric bending will eventually be required for EDM measurement using PTR, initially the bending
will be all-magnetic.

Magnetic fields are opposite in the two partial rings making up the figure-eight bending. 
Counter-circulating beams are not required.  Diametrically opposite bunches, such as ${\bf b2}$ and ${\bf b4}$,
collide at intersection point IP, at the center of the cross-over line.  

An important role of RF acceleration 
(which is not shown, and which necessarily averages to zero) is to phase lock the beam revolution frequency to an 
extremely accurate absolute frequency over long runs.  A possibly more important role is one which will 
require the quite high RF frequency required to prepare the short bunch lengths that will be
needed to achieve high luminosity.  If there is an ``Achilles's' heel'' to the present proposal, this is it.

The plan is to measure below-pion-threshold elastic $p,p$ (or $d,d$) scattering in 
(fixed-target-equivalent) $100<E<400$\,MeV laboratory energy range.  This would exploit the JEDI-Juelich-developed
\cite{CYR} long spin coherence time, spin phase-locked, pure spin-state, polarized beam technology.

Final-state, single-particle polarization measurements are sensitive to $T$-violation in double-spin 
observables.  For the goal of detecting T-violation in (measurably) elastic $p,p$ (or $d,d$) scattering, 
the detection chamber shown on the right in Fig.~\ref{fig:COSY-hall-mod6-racetrack} is located at the 
intersection point (IP) of the figure-8 ring.  

The proposed tracking chambers are up-down (or left-right as appropriate) symmetric, with nearly full $4\pi$ 
solid angle detection, except for the vertically-central section which is heavily compromised by the requirement 
that the colliding beams are crossing at right angles.  

Final state detection will be provided for nearly every elastic $p,p$ scatter, by stopping both scattered 
particles in coincidence in the up-down (or left-right) symmetric, polarization-sensitive, tracking chambers.
With nearly full acceptance, high-efficiency polarimetry is provided for every elastic $p,p$ scatter.

Description of this capability is especially simple since, unlike scattering 
from a target fixed in the laboratory, the storage ring frame of reference is close to the CM frame.  
The negligible spread of stopping energies makes it practical to study the full stopping
tracks of both scattered particles for every scatter.  This ``clean'' scattering detection provides a substantial 
advantage for colliding beam measurement compared to fixed target measurement.  Here ``clean'' particle 
detection means two unambiguous single tracks detected in time-coincidence, with accurately valid kinematics.  
Final state proton energies will be equal, which can be expected to provide sensitive elastic/inelastic 
selectivity.  These events will provide accurate spin-dependent differential cross section measurements 
over most of $4\pi$ steradians.  

\subsection{Luminosity estimation}

\ 

Colliding beam luminosity and collision rate estimates have been performed.  Based, as they are, on luminosity 
formulas that apply to head-on collisions, these formulas undoubtedly over-estimate the luminosity of right angle 
collisions.  From this perspective, our assumed data rates are over-estimates.  
\emph{Until realistic transfer line optics and realistic longitudinal beam dynamics has been established, 
the absolute rates used in this paper are not very reliable.}  One can say, alternatively, that the
run durations required to obtain the assumed numbers of scattering events is uncertain.

\subsection{Data rate estimation}

\ 

Other rate considerations are also important.  As well as enabling high beam polarizations, electron cooling inherited 
from COSY will reduce beam emittances, and energy spreads. From this perspective, our assumed rates are 
under-estimates.

With roughly one in 400 tracks scattering elastically in a carbon polarimetry tracking chamber, $10^6$ events 
are expected to exhibit at least one elastic $p$-carbon scatter.  With both incident beams
being, say, nearly 100\% up-polarized, these million events would be candidates for potential detection of spin flips 
forbidden by T-symmetry. These are favorable events in the sense that a substantial fraction are subject 
to polarimetric analyzing power averaging greater than, say, 80\%.  

\emph{Persuasive visualization of T-violation will be 
provided by unexpected correlation between the azimuthal scattering directions of 
coincident final state protons upon their entry to the tracking chambers, where the analyzing powers 
are close to 100\%.}
When all events are distributed by their (precisely-known) energies, their distributions 
in energy must match the dependence of analyzing power on energy, which is close to 100\% for tracks
starting close to the entrance to the detection chamber.  As a result the events carrying most of the statistical 
information are the ``early scatters'' occurring shortly after chamber entrance. This makes it sensible 
to tally up-down and left-right polarization determinations layer by layer in the tracking chambers.  
In other words, efficiencies and analyzing powers can be evaluated layer by layer, with events sorted 
into corresponding bins.

Of the $10^6$ events just discussed, roughly 2500 will show two clean polarization sensitive
scatters.  Events with both tracks showing carbon scatters close to the chamber entrance will be ``gold-plated'' 
in the sense that both scattered particle spin states have been measured with high analyzing power, 
meaning that their spin state, ``up'' or ``down'' is known with near certainty.  Of course, these 
will represent only a quite small fraction of the double-scatter events, but their statistical 
weight will be huge.
  
So far the distributions have been assumed to contain no T-violating scatters.  One sees, already,
the need for quite complicated data processing, in order to identify T-violating events. 
One therefore seeks initial spin state preparation that can be expected to maximize the 
fraction of events that are good candidates for counting as evidence of T- or 
P-violation\footnote{The discussion will continue to be ambivalent as to the separate
identification of T- and P-violation.  Since charge conjugation plays no obvious role in $p,p$
scattering, it is hard to avoid invoking PCT conservation, which would imply that any T-violation
has to be canceled by a P- violation.}.

Investigation to be spelled out next, has produced the initial spin state preparation shown in 
Fig.~\ref{fig:Correlated-spin-driven-pp-collider}.

\subsection{Forbidden ``null detection'' of T-violation}
Because of CPT symmetry, which is usually assumed to be sacrosanct, a violation of CP-symmetry is equivalent to 
a violation of T-symmetry.  Though CP violation of the weak nuclear force has been observed, 
this violation seems too weak to account for the imbalance of matter and anti-matter in 
the present day universe. Current models have both strong interaction and electromagnetic
interactions preserving P and T symmetry individually. Since charge conjugation plays no role in elastic 
nuclear scattering, CPT-conservation reduces to $PT$ violation for nuclear scattering, such as the elastic 
$p,p$ or $d,d$ scattering under discussion.  This does not make the detection of P-violation and T-violation
equivalent, however, for various reasons.   

Though weak (by definition) the weak nuclear force can be expected to play some role in elastic
$p,p$ or $d,d$ scattering. So the detection of P-violation, \emph{per se}, would be of less interest than
the detection of T-violation. More important, and considered next, is the fact that there are 
significant theory-based limitations to the extraction of T-violation from experimental measurement 
of scattering rates.

This issue was introduced by Arash, Moravcsik and Goldstein\cite{ArashMoravcsikGoldstein} who showed that,
independent of dynamical assumptions, no ``null T-violation experiment'' could be designed.  Here a 
``null experiment'' experiment is defined to mean any experiment for which a statistically significant 
non-vanishing observed counting rate would demonstrate time reversal violation. 

Before commencing their proof, these authors usefully define three classes of symmetry-violation detection 
experiment: (i) performing sequentially, an experiment followed by ``theoretically the same'' experiment
run backwards in time; (ii) measuring a ``self-conjugate'' reaction which, under time reversal, goes into
the same reaction; and (iii) a null experiment, as defined in the previous paragraph. 
The authors cite experiments of 
type (iii) that set fractional upper limits on parity-violation as small as $10^{-7}$. 
The simplest example of type (i) is the time-reversal scattering requirement for polarization P in a
forward scattering process to be equal to analyzing power A in the time-reversed process. 

Arash et al. point out the necessity of significantly different apparatus for type~(i)
counting rate comparisons, and (correctly, from this experimentalist's point of view) of an 
inevitable uncertainty in matching the acceptance apertures of two diverse experiments to an
accuracy very much better than 1 percent. It becomes increasingly difficult to reduce violation upper limits 
by an increasingly large factor.  

Arash et al. proceed to prove the impossibility of designing a  null experiment capable of detecting 
T-violation. Taken together, their arguments present a bleak future for direct experimental 
detection of T-violation.  This section reviews the Arash et al. results. A later section 
describes the extent to which our proposed storage ring experiment circumvents some of their 
detailed conclusions without contradicting their acknowledged validity. 

A predicate for the  Arash, Moravcsik and Goldstein (mathematically abstract) proof is that
measurable polarized state scattering rates are due exclusively to bi-linear combinations of scattering 
amplitudes. They do, however, acknowledge that ``there is one relationship that circumvents this 
constraint, \emph{namely the optical theorem}, (based on probability conservation) in which a rate
bi-linear in amplitudes is related to a rate linear in intensities, although only in a special way (namely, 
utilization of the real part of the forward reaction amplitude).'' 

Arash et al., also mention a proposal of Stodolsky\cite{Stodolsky} who had earlier pointed out the 
same ``loop-hole'' by suggesting an ``interesting time reversal test with polarized targets'' 
by measuring slow neutron-nuclei forward scattering. 

Quoting Stodolsky almost verbatim, for spin vectors of neutron and target respectively, he

\noindent
``imagines (beam polarizations) ${\bf s}_n$ and  ${\bf s}_T$ 
to be at right angles to each other, and to the \emph{beam direction} ${\bf{m}}$. Then, to restore 
the original  direction, he performs a $180^{\circ}$ around, say, the ${\bf s}_n$ axis, which 
reverses the direction of ${\bf s}_T$. Thus time-reversal invariance requires
\begin{equation}
F_{{\bf s_n}_x,{\bf s_T}_y } = F_{{\bf s_n}_x,{\bf-s_T}_y } 
\label{eq:StodolskyAmplitudes.1}
\end{equation}
where $x$ and $y$ are meant to show that  ${\bf s}_n$ and  ${\bf s}_T$ are oriented perpendicular to
each other, and to the beam axis $z$. In other words, detection of time-reversal violation favors
regions centered where
\vskip -0.5cm
\begin{equation}
F \sim ({\bf s}_n\times{\bf s}_T)\cdot {\bf{m}}\quad .
\label{eq:StodolskyAmplitudes.2}
\end{equation}
is maximal\footnote{Since the condition expressed in Eq.~(\ref{eq:StodolskyAmplitudes.2}) was 
derived assuming collinear incident beams, its application in the current situation is murky, 
other than suggesting that, for sensitivity to symmetry violation, nether spin direction should 
be parallel to either incident beam direction.}

Stodolsky continues by suggesting theoretically plausible sources of such a term in
the forward scattering amplitude.  Though useful for a neutral particle such as the neutron,
for charged particles, Rutherford scattering, peaking forward and back,
makes the forward nuclear scattering amplitude immeasurable for charged particles.
One cannot, therefore, consider using the optical theorem for $p,p$ scattering.   

\subsection{Detection chamber polarimeter properties}
Referring back to the Arash et al. categorization into experiment types, 
this proposed experimental configuration might seem to fit into the type~(ii) ``self-conjugate'' 
type---forward-time and reversed-time reactions being measured with identical apparatus. This 
is not precisely the case, however. The beam currents of bunch~1 and bunch~2 will not be 
exactly equal, and their spin directions reverse under time reversal. Neither will the apparatus be
exactly left-right nor exactly up-down mirror symmetric. 

Formally, the experimental apparatus is of category~(i)---it is not exactly self-conjugate.
But the apparatus is very nearly symmetric in most ways and there are beam-based
procedures that can greatly reduce normalization systematic uncertainties to well 
below levels values that Arash et al. suggest may be irreducible.  Furthermore, with nearly
full solid angle aperture coverage, rate determinations become branching ratio
determinations rather than cross section measurements.  

Whereas the ``denominator'' in a small aperture cross section measurement is a hard-to-determine 
solid angle, our branching ratio normalizing factor is a fixed total number of ``good'' events.  
Small solid angle apertures and near $4\pi$ solid angle denominators are both subject to systematic 
error, but the fractional error is much less for solid angles approaching $4\pi$.  
Furthermore, expressed as a branching ratio, the same denominator will be common to the 
two rates being compared.  Though the experiment is of category~(i) in the Arash labeling, the 
normalization uncertainty can become small relative to counting statistic errors.

\begin{figure}[h]
\centering
\includegraphics[scale=0.3090]{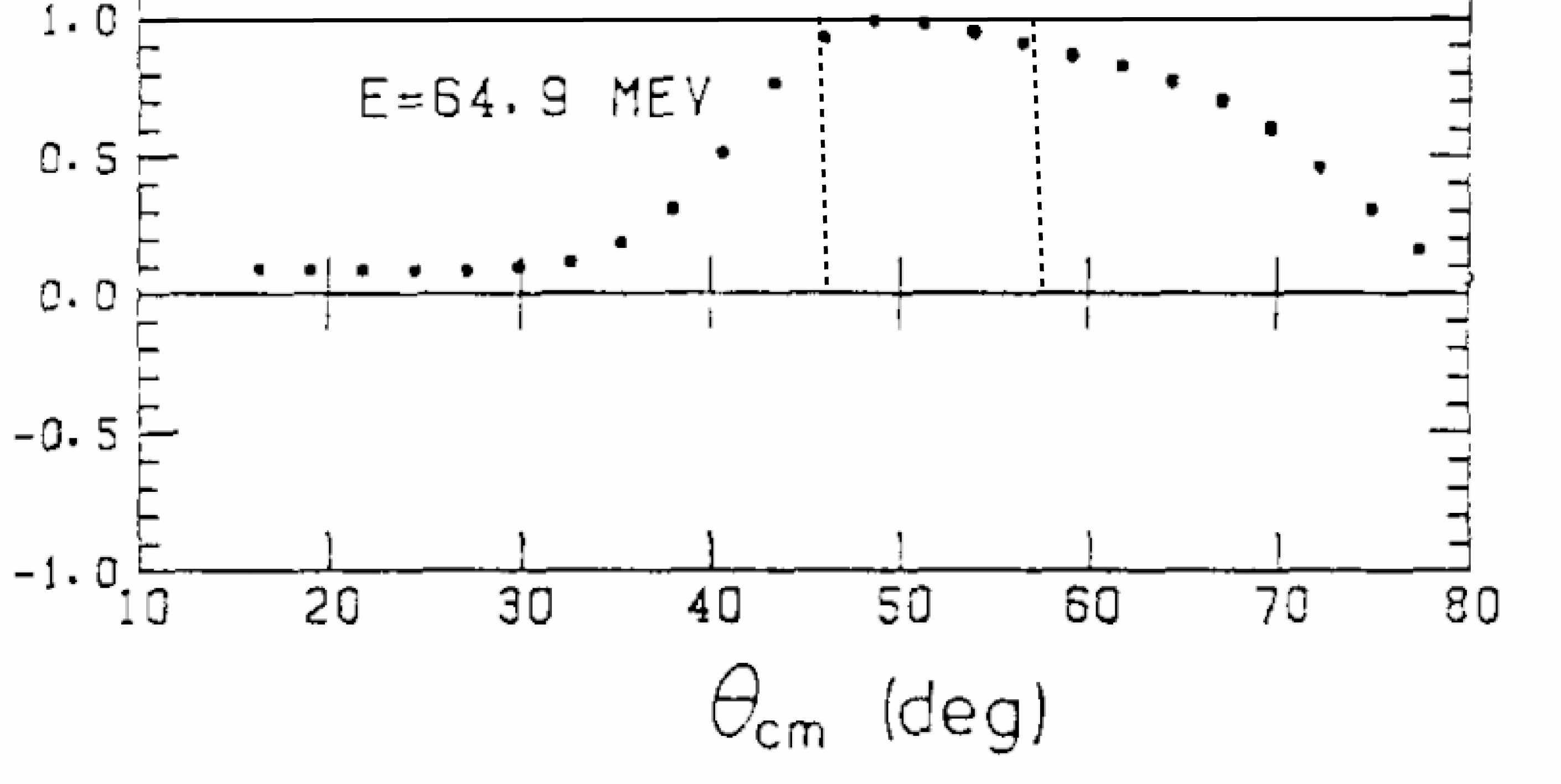}
\includegraphics[scale=0.355]{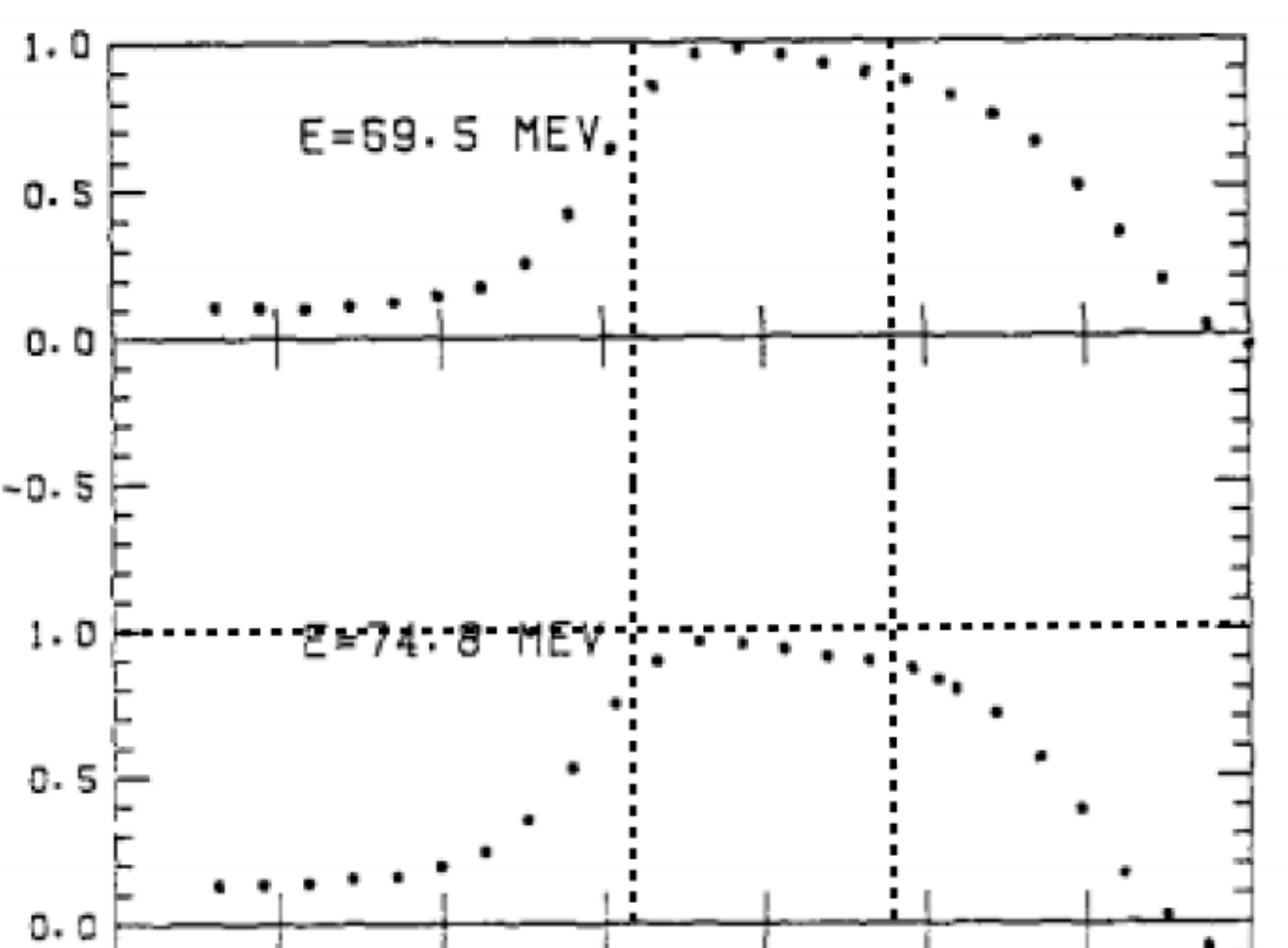}
\caption{\label{fig:p-C-analyzing-power}Center of mass energy
analyzing powers for left/right carbon-scattering
proton polarimetry\cite{Ieira}. At all energies shown the analyzing power peaks above 0.99
and remains higher than 0.9 in the angular range from (for example) 45 to 60 degrees, where
it can be accurately approximated by a polynomial fit tangent to the point at which
the efficiency is 100 percent.  With angular bins of one percent width in analyzing 
power, regions of predictably-high probability spin-flip violation can be
accurately delineated. Extended from 100${}^{\circ}$ to 170${}^{\circ}$, the plots are mirror 
symmetric about $\pi/2$. Since the angular distribution is known to be quite
accurately isotropic, this high sensitivity region represents roughly 1/4 of all events. Of 
these an elastic scatter occurs with high probability in the first 1/4 of the proton stopping 
range.  Especially valuable will be the one in 400 fraction of single scatters for which 
the polarizations of both scattered particle are measured with high analyzing power.}
\end{figure}

The tracking chamber plate thicknesses must be thin enough that no particles ``range out undetected'' 
even at the lowest particle energy.  Best foil thickness compromises have been worked out, especially, 
by Ieira et al.\cite{Ieira}. (A very small set of) their measured analyzing powers for scattered 
proton polarimetry are shown in Fig.~\ref{fig:p-C-analyzing-power}.  

As predicted by Plottner and Bacher\cite{Plottner-A-eq=1}, under not unlikely conditions
elastic scattering analyzing powers are guaranteed to be exactly 1 at some energy and direction 
in phase space.

Their conclusion is based on the understanding,  
for the scattering of initially unpolarized spin-$\frac{1}{2}$ particles from particles without spin,
that the polarization $P$ of the scattered beam can be expressed in terms of two complex quantities 
$f$ and $g$, the non-spin-flip and spin-flip components of the scattering amplitude, functions of
the scattering angle $0$ and of the energy $E$, as
\begin{equation}
P = \frac{2\Im(fg)}{f^2+g^2},
\label{eq:Polarization.1}
\end{equation}
where the amplitudes $f$ and $g$, can in turn be expressed in terms of partial wave
phase shifts.

Measurements of Przewoski et al.\cite{Przewoski} confirmed this for proton carbon-12
elastic scattering in the laboratory energy and angular ranges $160<T<200$\,Mev and 
$12^{\circ}<\theta_{\rm lab}^{\circ}$. In terms of $\Delta T = T - 183.1$\,MeV and 
$\Delta\theta = \theta - 17.75$\,degrees, their measurements were well fit by the formula 
\begin{equation}
 A(\theta,\Delta T) = 1 - \alpha{\Delta T}^2 - \beta{\Delta T}{\Delta\theta} - \gamma{\Delta\theta}^2,
\label{eq:A-eq1.0}
\end{equation}
\noindent
with\ coefficients  
\begin{align}
\alpha &= 1.21\pm0.07, \notag\\ 
\beta  &= 1.61\pm0.11, \label{eq:A-eq1.1} \\
\gamma &= 1.00\pm0.07, \notag
\end{align}
in matching units.

These results are applicable to carbon-12 in any form.  Graphite may be impractical as plate material for a 
tracking chamber; perhaps industrial multiplane graphene could be used. Graphite foil, such as used by 
Ieira\cite{Ieira} at much lower proton energies is probably inappropriate.
The graphite plates might have to be sandwiched between thin conducting foil, such as aluminum which, itself, 
provides significant polarization analyzing power.  The polarimetric properties of low-Z materials do not 
depend strongly on Z\cite{Bagdasarian}.

From existing measurements and models, with both initial spin states being known with near 
certainty, the left/right and up/down asymmetries of every scattered proton can be 
predicted with quite good accuracy.  Based on the one million primary scatters for which at
least one $p$-carbon scatter is observed, the dominant statistical uncertainty in dual-detected
scatters will come from fluctuations in the $p$-carbon scatter of the other proton.
\section{Unbalanced spin-flip T-violation detection \label{sec:SpinFlipStatistics.1}}
\subsection{The lore concerning elastic $p,p$ scattering\label{sec:NuclearLore}}
A definitive early review of low energy $p,p$ scattering was written in 1950, by J.D. Jackson and 
J.M. Blatt\cite{JacksonBlatt}, (here J\&B). Copying verbatim, their second paragraph reads

``The theoretical interpretation of proton-proton scattering experiments has been given in the classic papers
of Breit and collaborators' which not only constitute the pioneer work in this field, but also contain an exhaustive
treatment of the subject. For both neutron-proton and proton-proton scattering at low energies (below 4 Mev)
the de Broglie wave-length of the nuclear motion is large compared to the range of nuclear forces. 
Hence the nuclear interaction is effective only in the S-states (zero-orbital angular momentum) of the two-particle 
system.  (Experimentally, attributed almost entirely to S-wave scattering up to appreciably higher energies, of 
the order of 10 Mev.)  The exclusion principle limits the possible states of the proton-proton system. Two protons
in an S-state, in particular, have to have anti-parallel spins (be in the singlet spin state).  The triplet spin state
is forbidden. The S-wave proton-proton scattering thus involves only one nuclear phase shift, the one for 
the ${}^1$S-state.  In contrast to this, neutron-proton S-wave scattering involves two phase shifts, for the ${}^1S$-state
and ${}^3$S-states respectively.''

J\&B are implicitly applying a general ``exclusion principle'' here, to limit the possible states of the $p,p$ system.  
Without this, since the initial proton states are created incoherently, clearly the triplet spin state is \emph{not} forbidden.
Because the protons in both input and output states are indistinguishable, and (in the CM system) the protons are collinear, 
there can be no way to experimentally associate a particular input state proton with a particular output state proton.  
Though J\&B apply this principle only at a very low 4\,MeV energy, the same principle is presumably applicable at all 
energies.  However, the strict limitation implied for possible spin states is specific to the low 4\,MeV energy.
\emph{The application of this indistinguishability principle in Derbenev geometry is disputed in the present paper.}
Observed in the laboratory, for the vast majority of all collisions, there is no difficulty in associating unambiguously
each scattered proton with a particular incident beam.  

Fifteen years later, in 1965, in the third edition of their book, Mott and Massey\cite{MottMassey} 
repeat the same Jackson-Blatt argument concerning possible spin states and add as well: ``Analysis of the 
earliest observations which provided definite evidence of the nuclear scattering of protons in hydrogen 
showed that the non-Coulomb interaction between two protons cannot be very different from that between a 
neutron and a proton in in an $S^1$-state.''  It had been confirmed by then that, in fact, the non-electrical 
forces between nucleons are independent of the nucleonic charges---the so-called charge independence of 
nuclear forces.  Proposed initially by Heisenberg\cite{WignerEisenbud} before 1941,  
\emph{this isotopic spin conservation principle, has been confirmed by overwhelming evidence ever since,
and is accepted without dispute in the present paper.}

\subsection{Coarse integrated polarimetry averaging\label{sec:CoarseAveraging}}
There is a clear statistical advantage for triplet incident states in which particle spins
in both beams are all ``up'' or all ``down''.  In this case, there could 
be an overwhelming preservation of the same overload in the scattered states. This would improve the 
selectivity for detecting scatters with unmatched spin flips.  

But these are states for which T-violation is unlikely.  Previous discussion has 
suggested that T-violation (or P-violation) detectability favors spin configurations for which triple 
products formed from the spin vectors of incident or scattered particles, along with any
momentum vector, do not vanish.  For long spin coherence time (SCT) one favors incident conditions 
with predominant ``up'' (or ``down'') polarization for both incident bunches.  To maximize T-violation
sensitivity this vertical component is best augmented by mutually orthogonal transverse polarization 
components.

It is experimentally challenging to maintain beam polarization at fixed polar angle relative
to ``up'' or ``down''.  However, this requirement is similar to optimal conditions for EDM measurement, for which the 
spin remains horizontal, always pointing approximately forward or backward, with polar angle equal to $\pi/2$). 
Such beam polarization stability has been successfully achieved in COSY.\cite{CYR}.

The ``analyzing power'' $A$ in left/right scattering asymmetry for $N$ scattered particles
is defined by  
\begin{equation}
A = \frac{R-L}{R+L} = \frac{R-L}{N}, 
\label{eq:Polarimetry.1}
\end{equation}
where $R$, $L$, and $N=R+L$ are, respectively, the numbers of right, left, and total scatters.
Though $R$ and $L$ are stochastic quantities, subject to counting statistic uncertainty, it is 
usually legitimate to treat $N$ as the unambiguous total number of scatters.  In this case it 
is a branching ratio, rather than an absolute cross section, that is being determined. 
It is then natural to introduce normalized ``branching ratios'' satisfying the relations
\begin{equation}
p=n_R=R/N;\quad q=n_L=L/N; p+q=1,
\label{eq:p_and_q.1}
\end{equation}
where $p$ is the right-scatter probability and $q$ is the left-scatter probability.
Substitution of these equations into Eq.~(\ref{eq:Polarimetry.1}) produces
\begin{equation}
p = \frac{1+A}{2};\quad q = \frac{1-A}{2} .
\label{eq:p_and_q.2}
\end{equation}
On the other hand, the ``efficiency'', $E$, in scattering asymmetry covers the common
situation in which $R$ and $L$ are normalized by a number $N_{\rm incident}$ which is the total
number of particles passing thought the polarimeter, most of which register neither in the right
nor the left detector.  Then one defines efficiency $E$ as the fraction of incident particles
that actually register in the polarimeter; 
\begin{equation}
E = \frac{N}{N_{\rm incident}}.
\label{eq:p_and_q.3}
\end{equation}
In this case it is appropriate to treat $N$ as stochastic.

Usually, in polarization measurement,  there has to be a trade-off made between high efficiency, mediocre 
analyzing power, and low efficiency, high analyzing power.  What has typically been used in historic 
$p,p$ polarization measurements can be characterized as low efficiency, $E<0.0001$, and fair analyzing power, 
$A_{\rm min}\approx0.4$.  

Along with the high efficiency $E\approx1/400$ (the probability that an 183\,MeV
proton stopping in carbon will suffer an elastic nuclear scatter) the main design concept for the 
present proposal is to concentrate on detection of scatters for which polarimeter analyzing powers 
are as close as possible to 100\%.  In any case, best T-violation selectivity will require strong 
weighting of data from high analyzing power regions.

Selective detection of T-violating scatters depends critically on the analyzing power of the polarimetry,
which is, itself, dependent on scattered proton energy.  For best selectivity it is appropriate to
weight most heavily data in regions for which the analyzing power is highest. 

\subsection{Figure-eight luminosity penalty estimate}

\subsection{Anticipated data rates\label{sec:NearPerfectAnalyzingPower}}
Anticipated data rate performance can start with the calculated storage ring luminosity of 10 inverse millibarns per
second, capable of producing $N_0=2\times10^9$ clean scatters per year.
Of these events, a number $N_1=10^7$ will provide $p$-carbon polarimetry for one or other of the final state protons,
including $N_2=25,000$ for which both final state proton spins have been measured; with a fairly small fraction of
these referred to as ``gold-plated'', these events are ``silver-plated''.  All these rates are tabulated in 
Table~\ref{tbl:EventFractions}.  

Note, however, that in a certain sense, modulo polarimetric inefficiency, the kinematics, including spin,
of every one of the $N_1=10^7$ single polarimetric-detected events have been fully-determined.  This 
assumes time-reversal symmetry, along with the concession that time-reversal violation will occur 
only at the one percent level. One hopes this fraction will exceed one percent but cannot expect this
to reduce the sample size appreciably.  From these data rates it is not obvious which class of events holds the 
best statistical power for detecting T-violation.
\begin{table}[h]
\medskip
\centering
\begin{tabular}{cccccc}             \hline
 event class      &  symbol  & formula   & fraction    & symbol & events/year     \\ \hline
$p,p$ scatter      &  $N_0$   &    1     &     1       &  $N$   & $2\times10^9$    \\
single spin meas. &  $N_1$   &  $2/E$   &   2/400     &  $N_1$ & $1\times10^7$    \\
double spin meas. &  $N_2$   & $2/E^2$  &   2/$400^2$ &  $N_2$ & $0.25\times10^5$ \\ \hline
\end{tabular}
\caption{\label{tbl:EventFractions}Anticipated event rates with increasing detection quality per nominal
year running time for polarimetric detection efficiency $E=1/400$.}
\end{table}

Let us consider the experimental problem of detecting a correlation between 
initial and final spin coordinates, on the one hand, and scattering directions on
the other.  Based on many historic experiments, experimentally measured cross sections 
have been parameterized to best fit, PCTC (P and T-conserving) theoretical production
models. These distributions have been and can now be digitally reproduced with high accuracy by 
Monte-Carlo simulation, as parameterized by best fit partial wave expansion coefficients. 

One now speculates that, hidden in this data, is a PCTV, time-reversal violating contribution with 
differential cross section at perhaps the one percent level, which has so far evaded detection. 
We propose, therefore, to study 
the same process by direct measurement.  This might be called ``analog Monte-Carlo determination''.  
The reason we have to acknowledge the stochastic nature of the measurement, is that we have only beams with 
less than 100\% polarization and polarimetry with less than 100 percent analyzing power. 

A requirement of T-symmetry is that, if one spin flips, so also does the other. 
The PCTC (P-conserving, T-conserving) fit presumably satisfies a constraint $N_u = N_{\rm null}$ 
where $N_{\rm null}$ is the (dominant) theoretical rate, calculated from already-known ``measured cross sections'' 
with no apparatus equipment prejudice concerning the fractional contribution of T-violating or P-violating 
processes, but the theoretical expectation that these contributions have been smaller than some small value, such 
as one percent.

In our proposed experiment,
frozen-spin monochromatic protons in independently adjustable (almost) pure spin states collide inside a nearly $4\pi$
acceptance polarimeter.  The CM system motion in the lab is modest and scattered energies are identical.  Both 
scattered particles will stop in nearly full-acceptance tracking chambers. Some will undergo 
nuclear scatters in the tracking chamber plates. This will provides polarimetry, for example with efficiency 
$E\approx1/400$ and analyzing power $A\approx0.96$.  

At the cost of some duplication, to reduce confusion between scattering statistics and polarimetry
statistics, one can switch variables in formulas~(\ref{eq:Polarimetry.1}) through (\ref{eq:p_and_q.3}) 
from $R$ and $L$ to $N_u$ and $N_d$, which are the (unknown) numbers of ``up'' and ``down'' protons 
in a scattered beam.  In the proposed experiment this amounts to treating $p,p$ scattering as 
``virtual polarimetry'' for which the  ``up'' or ``down'' state of each scattered proton has not 
yet been detected. 

For convenient reference to conventional statistics notation we will, however, retain $p$ and $q$, satisfying
$p+q=1$, as the probabilities of events satisfying the binomial distribution.  And, for numerical example in
the  proposed experiment, the typical numerical values of analyzing power and efficiency are very different
from conventional polarimetry.

For example, we use the value $A=0.96$, which produces $p=n_u=N_u/N=0.98$ and $q=n_d=N_d/N=0.02$
with $p$ and $q$ fractional (i.e. summing to 1) binomial distribution probabilities.  As a binomial 
distribution, standard deviation about the mean (the same for both $N_u$ and $N_d$) is given by
\begin{equation}
\sigma_{\rm binomial} = \sqrt{Npq}\overset{e.g.}{\ =\ }0.14\,N^{1/2}.
\label{eq:Polarimetry.2}
\end{equation}
Quoted with error bars, relative to the null-expected rate, according to 
this standard deviation, the measured up-rate will be
\begin{equation}
N_u = pN_{\rm null}\Big(1 \pm \sqrt{\frac{pq}{N}}\Big)
     \overset{e.g.}{\ =\ } {\bf N_{u,measured}}\Big(1\pm\frac{0.14}{N^{1/2}}\Big),
\label{eq:Polarimetry.3}
\end{equation}
where (with ${\bf N_{u,measured}}$ shown bold-face for emphasis) the statistical significance of deviation 
from a null distribution mean can be quantified by its magnitude in units of the ``$\pm$error'' term.
The merit of this formulation in the present context is that the error can be calculated without the
need for having actual data. 

For the sake of definiteness, let us tentatively presume a current upper limit 
T-violating contribution of $\widehat\sigma^{\rm current\ T-ViolFrac} = 0.01$ exists, with error bars 
$\pm\sigma_{meas.}=\pm0.01$.  This would match the supposition that any higher fraction would have already 
been detected.  The experimental challenge will be to produce a T-violation signal with systematic error 
smaller than the counting statistics error.  We substitute $N=5\times10^6$ into Eq.~(\ref{eq:Polarimetry.3}) 
to produce an estimated fractional statistical error
\begin{align}
\sigma_{est.} = \frac{0.14}{2236} =& 0.63\times10^{-4},\hbox{or}\notag\\
\hbox{T-ViolFrac\ } =& 0.01\pm0.63\times10^{-4}.
\label{eq:Polarimetry.4}
\end{align}
Preliminary and vague as it is, this evaluation is given only to suggest that T-violation at significantly low
levels should be achievable.  

Averaged over all directions, this fractional error is likely to be  misleading
since the measured polarizations themselves may tend to cancel on average.  More granular estimations, based on
judicious preparation of the incident bunch polarizations, follow.

\subsection{Two particle T-violation detection\label{sec:HiEfficLowAnalyse}}
The binomial distribution, for $N_2$ trials, of ``success'' with probability $p$ and ``failure'' with probability 
$q$ is
\begin{equation}
P(x) = \frac{N_2!}{(N_2-x)!x!}\,p^xq^{N_2-x}.
\label{eq:binomial.1}
\end{equation}
%

It is instructive for understanding the logic, to start by calculating the probability
of correct determination for a single particle spin flip using a polarimeter with 
mediocre probabilities, such as $p=0.58$, $q=0.42$.
Already with 10 events, the average success probability exceeds 1/2. 
For data sets with $N_2$ increasing arbitrarily beyond 10, the success probability approaches 
$p=0.58$ and the failure probability approaches $q=0.42$. 

Supposing (with no justification) that protons from a 100\% ``up''-polarized beam produce 
only ``up''-polarized scatters, consider a scattered proton entering a polarimeter for which 
$p=0.58$ represents the probability of its spin state being properly identified.
Further supposing T-violation consisted only of events with one spin flipping and one not, 
$p$ and $q$ could be described as ``null probabilities'' in the sense that, with no T-violation, 
the true distribution has no spin flips.  In the large sample limit a collected data set 
would then be expected to center on $x=0.58$, because this is the fraction of correctly 
identified polarizations.  
 
Contemplating the presence of some T-violating scatters, we would interpret deviation 
of the peak position from x=0.58 as indication of T-violation.  Regrettably, we do not actually ``
understand the probabilities''.  The true analyzing power will not, in fact, be constant, 
independent of kinematic parameters, nor will it even be exactly equal to $A=0.4$ ``on the average''.  
Though ameliorated by calibration efforts, this will inevitably leave detection of T-violation, 
purely on the basis of deviation from a calculated value (such as 0.58) as unreliable.

In conventional language, for small samples, the uncertainty will be dominated by 
statistical uncertainty but, for large samples, the uncertainty will be dominated
by systematic errors resulting from our imperfect understanding of the apparatus.

To reduce this systematic error it is important for the polarimeter analyzing power to
be as high as possible---at 100\% there would be no systematic error.  In the proposed
experiment, only a tiny fraction of the detected scatters will have analyzing power greater
than 0.99, but a substantial fraction will have a, still respectable, analyzing power 
greater than 0.9.

In an experiment intending to investigate time reversal conservation the distinguishing signal for 
T-violation is that, if one spin flips, so also must the other.   If spin-flips were forbidden 
by T-conservation (\emph{which they are not}) then any detected spin-flip would provide evidence
for T-violation.  To reduce confusion it is sensible for initial spin states to be set up
to be all the same, say ``up''.  Then any detected ``down'' polarized scattered proton would
provide evidence of T-violation.  Since \emph{some T-conserving amplitudes do involve spin-flips},
the true situation would be much more complicated, except for the fact that, for T-conservation
in elastic scattering if one spin flips, so must the other.  T-conserving scatters must
``match'' in this sense.

As a consequence, except for different interpretations and values for the $p$ and $q$ probabilities, 
the  T-violation probabilities associated with correctly identifying T-violation using ``mismatched'' 
spin flips is described by the same binomial distribution, Eq.~(\ref{eq:binomial.1}), as for the single 
particle T-violation described previously.

The least ambiguous signature for T-violation is one spin flipping, and the other not. 
Actually, both identifications being wrong provides information equivalent to both being correct, 
as far as T-violation detection is concerned.  As a result, in a 98\%  analyzing power set-up,
the two particle correct T-violation identification probability is: $0.98\times0.98+0.02\times0.02=0.9608$. 
The probability of one correct and one incorrect identification is $2\times0.02\times0.98=0.0392$.  
Conservation of probability, confirmed by summing the probabilities, is reassuring.

Let us assume $N_1=5\times10^6$ events, for which all kinematic parameters are known and at least 
one scattered particle's polarization, have been recorded.  For each event there is also a recorded 
``phase angle'' $\psi$ providing the instantaneous orientations of both input spin vectors in spin-space.  
For now this is just a tallying mechanism for associating the instantaneous polarization 
measurement with the other particle parameters in effect when the polarization measurement was made.
Each of the measured polarization values can then be compared with the previously established 
``null polarization values''.

Of the events just discussed, about $N_2\approx0.25\times10^5$ events, will contain
spin information about both scattered particles.  Some, with almost 100 percent analyzing power, 
were referred to earlier as ``gold-plated''.  Even a few events in which one spin flipped, and one did not, 
would prove the existence of T-violation.  Further statistical analysis of these events is far too difficult
to be attempted in the present paper.

\section{Application of ``spin transparency''}
\subsection{History}
The concept of ``spin transparency'', was introduced and defined in 2016 by Derbenev\cite{Derbenev}: 
``In a figure-8 collider, the spin first rotates about the 
vertical field in one arc.  This rotation is then undone by the opposite field in the other arc. 
The resulting effect of the “strong” arc dipoles on the spin dynamics reduces to zero over one
particle turn and the whole ring becomes “transparent” for the spin.'' The concept is more fully 
developed and explained in Filatov et al.\ \cite{Filatov} 
In the context of the present paper, it is spin transparency (which is unrelated to ``spin filtering'') 
that contributes to nuclear spin physics by providing the accelerator physics capability to study 
T-violation in low energy nuclear physics.

A property that makes the term ``transparency'' apt, is that, regarded as an approximate method,
the transparency approximation can usually be most valid for slow changes on long (adiabatic approximation) 
time scales.  The next section shows how adiabatic sinusoidal variation of beam spin states can provide 
Fourier sensitivity enhancement to enable experimental detection of T-violation in $p,p$ scattering.  

\subsection{Adiabatic Fourier sensitivity enhancement\label{sec:Fourier}}
"Can One Hear the Shape of a Drum?" is the title of a 1966 article by Mark Kac in the American 
Mathematical Monthly which made this question famous.  Not having read the article (since the title
expresses a question rather than an answer) one can ask a similar question in the present context: 
``With the proton assumed to be composite, rather than elementary, can internal proton variability 
have a measurable effect on ``elastic'' $p,p$ scattering?''.     

A routine procedure of experimental mechanical engineering is to scan in frequency a harmlessly small 
sinusoidal signal applied judiciously to a physical structure, in order to determine
frequencies at which strong drive might be harmful.  Applying the principle that any linear response
must have the same frequency as the drive, synchronous detection can enhance the sensitivity by
a large factor. 

Here, for elastic proton scattering, we conjecture that the answer to the rhetorical question above is ``yes''.  
A collision between two protons could temporarily stir up the contents of the protons enough to affect the 
elastic angular scattering distribution while producing no other detectable particles nor any measurably-large 
reduction of proton energy.  

\begin{figure*}[htb]
\centering
\includegraphics[scale=0.14]{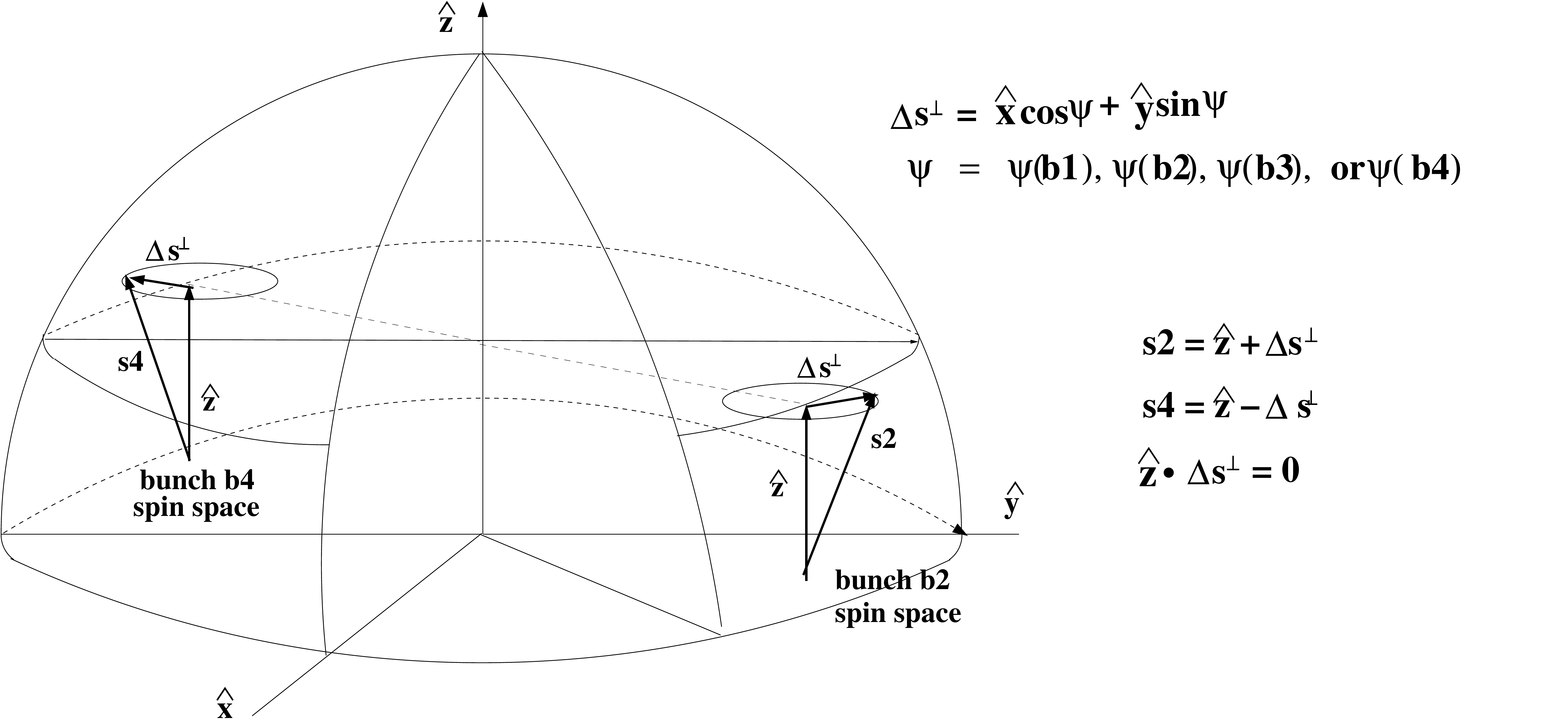}
\includegraphics[scale=0.14]{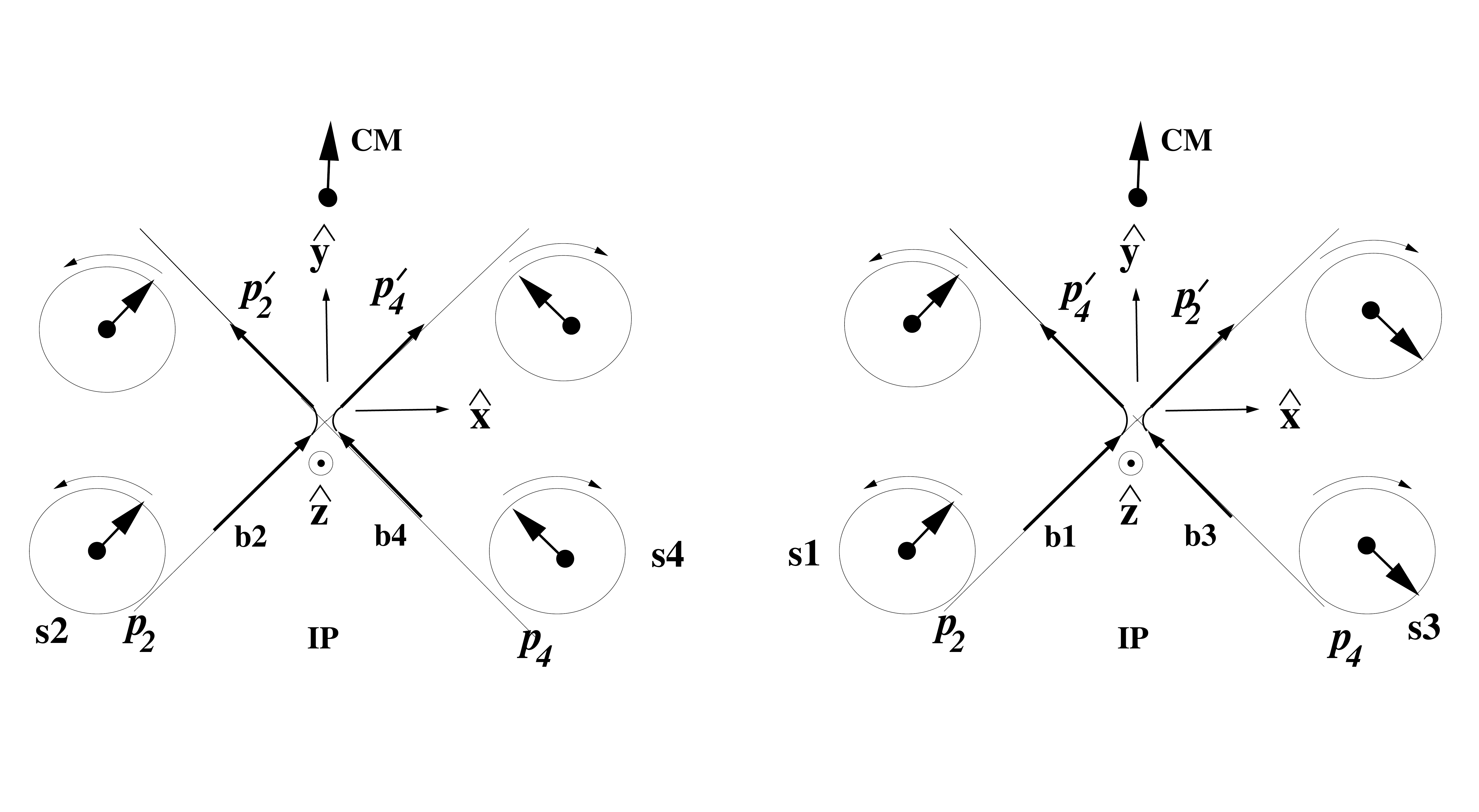}
\caption{\label{fig:Correlated-spin-driven-pp-collider}Exploiting adiabatic spin transparency, beam spin orientations
can be scanned coherently, for example to achieve Fourier series statistical enhancement of T-violation detection sensitivity; 
see Section~\ref{sec:Fourier}.
{\bf Left:\ }Two of four beam bunch spins are shown. Colliding pairs are ${\bf b1}$,${\bf b3}$ and ${\bf b2}$,${\bf b4}$.
Phase angles $\psi$ are externally controllable independently for each of the four beam bunches.
{\bf Right:\ }Spins of bunches ${\bf b1}$ and ${\bf b2}$ are the same, ${\bf b3}$ and ${\bf b4}$ opposite.
Preparation of two or more deuteron bunches in individualized spin states as assumed in this figure, 
could be demonstrated immediately at COSY, using the methods described in reference~\cite{RathmannWienFilter}.
}
\end{figure*}

A phase angle $\psi$ is introduced in Fig.~\ref{fig:Correlated-spin-driven-pp-collider} with the
considerations just introduced in mind.  A way suggested to enhance the statistical sensitivity to
T-violation is to vary $\psi$ sinusoidally and detect the polarimetric response synchronously.

Fourier series capability is especially valuable for statistically ``noisy'' data.  In our case 
the statistical power of individual events varies over such a large range that most of the events provide 
little help in studying spin dependence.  A bias-free strategy for filtering
promising events from useless events is needed.  With relatively few significant Fourier coefficients being
expected, for example three, constant term plus fundamental harmonics, it is useful to allow every event 
to contribute to every coefficient, with the expectation that the contributions from especially noisy 
events will average to zero.

This Fourier series capability should be especially useful for analyzing the $N_1=5\times10^6$ 
``silver-plated'' events for which the spin of just one final state proton has been measured. 
The data sets will be periodic functions of the $\psi$-phase angles.
With the data represented as a periodic function, with period $2\pi$, it is natural to
convert the data sets into Fourier series.  The proton beam polarization could advance 
nominally (for example) as ${\bf s_p} = \cos\psi\,{\bf\hat x} + \sin\psi\,{\bf\hat y}$,
%
%
but with individual bunch phases separately adjusted.  The T-violation evidence will then 
be contained primarily in the constant term plus the lowest harmonic sine and cosine 
coefficients of the Fourier series expansions of the data sets.

The uncertainties in these T-violation determinations should then be dominated
by statistical errors.  The systematic error should be quite small, since the 
data will be ``self-normalizing''; the normalization will be established by the constant 
term of the Fourier series. Presuming that the double-polarization 
data is more statistically significant, the single particle measured polarization 
data, should provide a self-consistency check.

One can also consider restricting the analysis to regions for which the polarimetry analyzing 
power is well above average.  A significant fraction of $p,p$ scatters will have laboratory 
scattering angles corresponding to the central CM angular ranges 
shown bounded by broken lines in Fig.~\ref{fig:p-C-analyzing-power}. In this range 
the polarimetry analyzing power starts above 97\%. Furthermore, 
mentally-interpolating from the upper and lower plots in Fig.~\ref{fig:p-C-analyzing-power}, the 
analyzing power will exceed, for example, 70\%, for a substantial fraction of their full-ranges in the 
tracking chamber. 

\section{Recapitulation and acknowledgments}
The thesis of the present paper is that $p,p$ scattering near the pion production threshold 
represents the most promising region for the investigation of wave particle duality, by 
experiment and theory.

The regions of validity of ``correct'' theories in physics are always bounded.  Certainly the region
under discussion represents a boundary beyond which classical mechanics ceases to be valid.  
Alternatively, this region represents a boundary below which quantum physics becomes unnecessary. 
Here the word ``quantum'' is ambiguous.  As introduced by Planck, the term quantum applied to 
massless ``particles''.  As applied in elementary particle physics, the term has come to apply 
also to massive unstable particles. 

Without a doubt, the Einstein formula $\mathcal{E}=mc^2$ corrects Newtonian mechanics for massive 
particle speeds that are not negligibly small compared to the speed of light.  But, when the ``mass'' 
of an unstable particle comes into question, so also does the Einstein formula.  To good approximation, 
this issue is addressed by the Heisenberg time uncertainty principle; but this may represent only a 
temporary ``fix'' for the Einstein formula and quantum mechanics.  The $p,p$ region under discussion is most appropriate
for investigating this issue experimentally.

The issue of angular momentum is more serious.  There seems to be some kind of trade-off, among
mass, spin,  and angular momentum.  Surely, quantum mechanics, as augmented by the 
incorporation of spin 1/2 by Pauli and Dirac, provides an unquestionably valid and accurate theory 
of atomic physics.  But the $p,p$ scattering region under discussion calls these ideas into question
for nuclear physics. And ``Why does the muon 'weigh?''' continues to demand some kind of answer.

The ``center of mass'' concept is central to quantum mechanics, but foreign to special relativity,
where it is a mathematical abstraction, of no instantaneous geometric significance.  The upper and 
lower pairs of Dirac spinor amplitudes could just as well represent fractional negative and positive mass 
fractions in a classical sense.  The muon (like the electron, but unstable) could play the 
transitory role of smoothing over the kinks and distortions in the transmutation from
two $p,p$ to three $p,p,\pi$ ``particles'' with the $\pi$ itselve unstable, and the neutrino 
invisibly cleaning up the resulting mess.

This could very well cause the outcumes in parallel incidence and perpendicular incidence to
be strikingly different.  The proposed apparatus could demonstrate such differences, In glancing
collisions in DERBENEV geometry, the geometric distortions in glancing collisions can be more
easily visualized than in WOLFENSTEIN geometry.   
 
This paper has shown, in Sections~\ref{sec:RoleOfSpin} through \ref{sec:AnomMDMs}, 
that correct relativistic treatment of the ``anomalous'' magnetic moment of 
the proton explicitly contradicts $p,p$ time reversal invariance by the unavoidable introduction of 
spin precession in (weak, but unavoidable) magnetic field encountered during scattering through 
a finite angle.  It also suggests, for example in Figure~\ref{fig:EikonalTrajectories}, a treatment 
of scattering from spheres resembling that of Van de Hulst\cite{Van-de-Hulst}.
The near coincidence between deBroglie wavelength and proton radius, shown in 
Figure~\ref{fig:deBroglie-lambda-vs-E_tot}, and pion threshold, shown in 
Figure~\ref{fig:pp-total-crsctn-vs-KE}, can hardly be accidental.  

Almost a century after the beginning of nuclear physics (in Europe, mainly German-speaking, by Heisenberg, Wigner, 
Fermi, Bethe, Chadwick, and others) $p,p$ scattering is described entirely by phenomenological partial wave analysis, 
which makes very little sense either experimentally or from any fundamental theoretical perspective.  The footnote 
below, copied verbatim from page~415 of Stratton's book\cite{Stratton},
\footnote{The first thorough investigations of the electromagnetic problem of the sphere were made by Mie,
\emph{Ann. Physik,}, {\bf 25}, 377,1908, and Debye, \emph{Ann. Physik,}, {\bf 30}, 57, 1909.
Both writers made use of a pair of potential functions leading directly to our vectors {\bf M} and
{\bf N}. The connection between these solutions and a radial Hertzian vector has been pointed out by Somerfeld
in Riemann-Weber, Differentialgleichungen der Physik, p. 407, 1927.} 
anticipated the problem.

Neither the book by Van de Hulst\cite{Van-de-Hulst} nor by Stratton ever mention the term ``angular momentum'' 
nor partial wave expansion, while, at the same time, describing a formalism more nearly matching observable 
characteristics of $p,p$ scattering.  

The history of modern physics can be encapsulated in a few sentences (cut and pasted from the web):
Johannes Kepler enunciated his laws of planetary motion in 1618.  Galileo officially faced the Roman Inquisition 
in April of 1633 and agreed to plead guilty in exchange for a lighter sentence.  Newton unified Galileo's theory 
of falling bodies with Kepler's laws of planetary motion. He published his laws of motion and universal gravitation 
in 1687.  Albert Einstein's 1905 theory of special relativity is one of the most important papers ever published 
in the field of physics.  Bill number 246, debated by the Indiana General Assembly in 1897, proposed establishing 
by law the value of $\pi$ to be 3.2. The bill never became law due to the intervention of Prof. C. A. Waldo 
of Purdue University.  And so on, as described above in this paper.

In its wisdom, and not to be outdone by the state of Indiana, the German government has determined that COSY has 
outlived its usefulness. This paper has attempted to make the case that this action rivals the Roman Inquisition 
in stupidity.  The correct description of power generation by nuclear fusion relies on the correct understanding 
of nuclear transmutation, which can only proceed from a correct understanding of nucleon, nucleon scattering.
Certainly the cost of the program proposed in this paper is negligible in comparison to the cost and necessity 
of the development of power generation by nuclear fusion.

I wish to acknowledge the valuable hints, corrections, or enthusiasm, by Kurt Aulenbacher, 
Yunhai Cai, Slava Derbenev, Eanna Flanagan, Susan Gardner, Joe Grames, Kolya Nikoliev, 
Maxim Perelstein, Frank Rathmann, James Ritter, Joken Stein, Saul Teukolsky, to my son John and 
other COSY collaborators, and to discussions concerning the frontier between mathematics and physics 
during regular weekly meetings with Leonard Gross.

\appendix

\section{Consistent treatment of $g$,$G$ and $g\rightarrow G$\label{sec:MDMs}}
\subsection{Orders of magnitude\label{sec:Orders-of-magnitude}}
A chronological listing for the values of nuclear physical constants includes, for example, (2014 values), (2010 values) 
(2006 values) (2002 values) (1998 values) (1986 values) (1973 values) (1969 values)\cite{NIST-background}
For example,  ``Atomic Weights and Isotopic Compositions for All Elements''.\cite{NIST-isotope-abundance}

The following statements, all expressed in MKS units, can be made about protons, electrons, and
protons, as they enter this paper:
\begin{enumerate}
\item
The radius of the proton is about 1\,fm ($10^{-15}$\,m), smaller than the radius of a hydrogen 
atom by a factor of 60,000. 
\item
The ``classical radius'' of a particle is given by $r^{\rm class.}= e^2/(4\pi\epsilon_0 mc^2),$ where 
$e$ and $m$ are the charge and mass of the particle.  This is the radius a charge having mass 
$m$ would have if its mass (uniformly distributed) and energy (derived according the E\&M) were
related by the Einstein formula. 
\item
The classical radius of the electron is given by $r_e^{\rm class.} = 2.8179403262\times10^{-15} {\rm m}$,
roughly three times the measured proton radius $r_p^{\rm meas.}\approx 1$\,F (for Fermi), but immeasurably
greater than the electrons actual radius, since the electron is considered to be and
treated to be ``point-like''.  (Mnemonic: if the (spherical) proton's interior had its own charge and an 
electron's charge superimposed continuously and uniformly, the overall system would be uncharged''.)
\item
The de-Broglie wavelength of a non-relativistic particle of mass $m$ and kinetic energy $K$
satisfies the equations $\lambda = h/p$,\ $K = p^2/(2m)$,\  $p = \sqrt{2mK}$.   Applied to 
electrons and protons;
$$ \lambda^{\rm de-B}_e \approx h/\sqrt{2m_eK_e},\quad \lambda^{\rm de-B}_p \approx h/\sqrt{2m_pK_p}.$$
$$\frac{\lambda^{\rm de-B}_e}{ \lambda^{\rm de-B}_p} = \sqrt{\frac{m_e}{m_p}\,\frac{K_e}{K_p}}$$
which, for comparison of electron and proton inelastic scattering  is approximately 
$$\sqrt{\frac{1}{2000}\, \frac{20\,{\rm eV}}{100\,{\rm MeV}}} \approx 10^{-5}$$
\item
For an electron with $KE = 1$\, eV and rest mass energy 0.511\,MeV, the associated DeBroglie wavelength is 1.23\,nm, 
about a thousand times smaller than the wavelength of a 1 eV photon; a 1 eV photon has a wavelength of 1240\,nm, 
with negligible uncertainty.
\item
The Compton wavelength of a particle is given by $\lambda^{\rm Compton} = h m c$.  
The Compton wavelength of the electron is $2.42631023867(73)×10^{-12}$\,m,
2000 times smaller than the proton Compton wavelength, which is therefore about $10^{-9}$\,m.
\item
The kinetic energy threshold for pion production in proton-proton collisions is 280 MeV, 
\item
These numerical examples serve as reminders or the huge differences in orders of magnitude of potentially
relevant physical parameters. The near coincidence between deBroglie wavelength and
proton radius, shown in Figure~\ref{fig:deBroglie-lambda-vs-E_tot}, and pion threshold, shown in 
Figure~\ref{fig:pp-total-crsctn-vs-KE}, can hardly be accidental.  
\end{enumerate}

\subsection{Integer arithmetic, $g\rightarrow G$}
The data table below exhibits  $g\rightarrow G$ conversion using of integer arithmetic---this issue
applies only to values of $m$ in the ``Atomic mass'' column, from NIST tables\cite{NIST-background}\cite{NIST-isotope-abundance}; 
integer entries have no decimal points and integer arithmetic is used in the evaluation of $G=(g\times m/Z-2)/2$. 
The motivation for this is that, along with the obvious requirement that the values of $m$ have to be identical in the 
evaluation of $g$ and $G$, there are no transcendental numbers in the relation between $g$ and $G$.  The importance
of respecting this discipline is apparent in many examples in the table; powers of ten, indicated by exponential notation 
appearing in $m$ column, show up as corresponding trailing zero digits in $G$ denominators.

The first task is to best approximate $g$ as a ratio of integers $g\approx n/d$, with denominator $d<100$, plus a correction
provided by the most accurate value of $m$ available.  The maximum value 100, was chosen such that at least the first 
four digits of the $g$ correction were zero---meaning accuracy, roughly speaking, to better than one part in ten thousand..  

Because the deuteron to proton mass ratio is known to such high accuracy three determination
are given; the surprising $g-G=1$ at low accuracy is seen, at higher accuracy, to be simple coincidence.  
In most other cases two evaluations are given.  In the cases of lithium, boron, and silver, two stable isotopes are treated.
This provides warning that the purity of samples needs to be taken into account.  

Other than noting that all entries have 
been taken from NIST tables, the sources of entries and their errors are not given in this paper; in fact most values of $m$ 
simply been truncated to seven digits, to reduce confusing complexity.  In every case the three lines providing 
``Relative Atomic Mass'', ``Isotopic Composition'', and ``Standard Atomic Weight''have been cut and pasted from NIST 
tables, \cite{NIST-background}\cite{NIST-isotope-abundance}---simple detective work can then identify most of the sources.  
As far as I know there are no reliable tables providing accurate values of $G$.

\begin{widetext}
\scriptsize
\begin{verbatim}\
%---------------------------------------------------------------------------------------------------------------------------------
% first period                                                 m           G=(g-fac/Z*m-2)/2      
%---------------------------------------------------------------------------------------------------------------------------------
 Z A -/*   element    S  g-fac(meas.)  ratio  correction    Atomic-Mass         G-(rational)         G        ratio    correction 
%---------------------------------------------------------------------------------------------------------------------------------
 1 1   H   hydrogen  0.5 +5.58569468 +391/70  -0.0000196.  1000000/10^6           251/140        +1.79285714 +147/82   +0.00017420
 1 1   H   hydrogen  0.5 +5.58569468 +391/70  -0.0000196   1007825/10^6       1451769/800000     +1.81471125  +49/27   -0.00010356
%                                     Relative Atomic Mass = 1.00782503223
%                                     Isotopic Composition = 0.999885(70)
%                                   Standard Atomic Weight = [1.00784,1.00811]
 1 2   H   deuteron  1.0 +0.85743820    +6/7  +0.0002953 1998463/10^6         -1004611/7000000    -0.14351585   -1/7   -0.00065870
 1 2   H   deuteron  1.0 +0.85743820    +6/7  +0.0002953 2014101/10^6         -957697/70/10^6      -0.13681385 -13/95  -0.00002825
 1 2   H   deuteron  1.0 +0.85743820    +6/7  +0.0002953 2014101778/10^9   -478847333/35/10^8      -0.13681352  -13/95  -0.0000282
 %                                Relative Atomic Mass = 2.01410177812
%                                     Isotopic Composition = 0.000115
%                                   Standard Atomic Weight = [1.00784,1.00811]
 1 3 * H   triton    0.5 +5.95799369 +566/95  +0.0000989   2992631/10^6     751914573/95000000   +7.91489024 +372/47   +0.00000337
 1 3 * H   triton    0.5 +5.95799369 +566/95  +0.0000989   3016049/10^6           566/95         +7.98465123 +519/65   +0.00003584
%                                     Relative Atomic Mass = 3.0160492779
%                                    Isotopic Composition = 
%                                    Standard Atomic Weight = [1.00784,1.00811]
 2 3   He  helion    0.5 -4.25499544 -217/51  -0.0000934   2992612/10^6     -12549953/3000000    -4.18331767 -251/60   +0.00001566
 2 3   He  helion    0.5 -4.25499544 -217/51  -0.0000934   3016029/10^6    -286159431/68000000   -4.20822692 -101/24   +0.00010641
%                                    Relative Atomic Mass = 3.0160293201
%                                    Isotopic Composition = 0.00000134
%                                    Standard Atomic Weight = 4.002602
 2 4   He  alpha     0.0  0.0                              3971525/10^6   
%                                    Relative Atomic Mass = 4.00260325413
%                                    Isotopic Composition = 0.99999866
%                                    Standard Atomic Weight = 4.002602
%--------------------------------------------------------------------------------------------------------------------------------
% second period                                                 m           G=(g-fac/Z*m-2)/2 
Z A -/*   element    S  g-fac(meas.)  ratio   correction    Atomic-Mass         G-(rational)         G        ratio    correction      
%--------------------------------------------------------------------------------------------------------------------------------
 3 6   Li  lithium   1.0 +0.8220473   +60/73  +0.0001294   5968419/10^6      -1331581/7300000   -0.18240836   -17/93  +0.000387330    
 3 6   Li  lithium   1.0 +0.8220473   +60/73  +0.0001294   6015122/10^6       -642439/3650000   -0.17601068   -16/91  -0.000186504
%                                    Relative Atomic Mass = 6.0151228874
%                                    Isotopic Composition = 0.0759(4)
%                                    Standard Atomic Weight = [6.938,6.997]
 3 7   Li  lithium   1.5 +2.170951   +165/76  -0.0001016   6961529/10^6     +46176819/30400000  +1.51897431  +120/79   -0.00001303
 3 7   Li  lithium   1.5 +2.170951   +165/76  -0.0001016   7016003/10^6     +46776033/30400000  +1.53868529   +20/13   +0.00022375
%                                    Relative Atomic Mass = 7.0160034366
%                                    Isotopic Composition = 0.9241(4)
%                                    Standard Atomic Weight = [6.938,6.997]
 4 9   Be  beryllium 1.5 -0.78495     -73/93  -0.0000037   8942209/10^6   -1396781257/744000000 -1.87739416   -92/49   +0.00015686
 4 9   Be  beryllium 1.5 -0.78495     -73/93  -0.0000037   9012183/10^6    -467296453/248000000  -1.88425989 -179/95   -0.00004936
%                                    Relative Atomic Mass = 9.012183065(82)
%                                    Isotopic Composition = 1
%                                    Standard Atomic Weight = 9.0121831
 5 10   B  boron     3.0 +0.600215      +3/5  +0.0002150   9935193/10^6     -20194421/50000000 -0.40388842    -21/52   -0.00004226
 5 10   B  boron     3.0 +0.600215      +3/5  +0.0002150  10012936/10^6      -2495149/6250000  -0.39922384     -2/5    +0.00077616
%                                    Relative Atomic Mass = 10.01293695
%                                    Isotopic Composition = 0.199(7)
%                                    Standard Atomic Weight = [10.806,10.821]
 5 11   B  boron     1.5 +1.7924326   +95/53  -0.0000202  10923826/10^6      50776347/53000000  +0.95804428   +91/95  +0.000149543
 5 11   B  boron     1.5 +1.7924326   +95/53  -0.0000202  11009305/10^6      20635359/21200000  +0.97336599   +73/75  +0.000032656
%                                    Relative Atomic Mass = 11.00930536
%                                    Isotopic Composition = 0.801(7)
%                                    Standard Atomic Weight = [10.806,10.821]
 6 12   C  carbon    0.0  0.0                             11906828/10^6    
%                                    Relative Atomic Mass = 12.0000000
%                                    Isotopic Composition = 0.9893(8)
%                                    Standard Atomic Weight = [12.0096,12.0116]
 6 13   C  carbon    0.5 +1.4048236   +59/42  +0.0000616  12902393/10^6      36748741/72000000   0.51039918   +49/96   -0.00001748 
%                                    Relative Atomic Mass = 13.00335483507
%                                    Isotopic Composition = 0.0107(8)    
%                                    Standard Atomic Weight = [12.0096,12.0116]
 6 14 * C  carbon    0.0  0.0                             13894516/10^6                             
%                                    Relative Atomic Mass = 14.0032419884
%                                    Isotopic Composition = 
%                                    Standard Atomic Weight = [12.0096,12.0116]
 8 17   O  oxygen    2.5 -0.757516    -25/33  +0.000058  16867145/10^6      -7597429/4224000   -1.79863376  -178/99  -0.00065396
%                                    Relative Atomic Mass = 15.99491461957)
%                                    Isotopic Composition = 0.99757(16)
%                                     Standard Atomic Weight = [15.99903,15.99977]
 9 19   F  fluorine  0.5 +5.257736   +510/97  +0.0000040  18850894/10^6     131132599/29100000   4.50627488  +356/79  -0.0000542
%                                     Relative Atomic Mass = 18.99840316273
%                                     Isotopic Composition = 1
%                                     Standard Atomic Weight = 18.998403163
 10 21   Ne neon      1.5 -0.441198    -15/34  -0.0000215 19992440/10^6      -4899433/3400000   -1.44100971  -134/93  -0.0001494
%                                     Relative Atomic Mass = 19.9924401762
%                                     Isotopic Composition = 0.9048
%                                     Standard Atomic Weight = 20.1797
%------------------------------------------------------------------------------------------------------------------------------ 
% fifth period, second row transition metals                    m           G=(g-fac/Z*m-2)/2 
Z A -/*   element    S  g-fac(meas.)  ratio   correction   Atomic-Mass         G-(rational)         G        ratio    correction
%------------------------------------------------------------------------------------------------------------------------------- 
47 107  Ag silver    0.5 -0.22714      -5/22  +0.0001327  106905091/10^6   -520505091/4136000   -1.25847459  -112/89  -0.0000476
%                                      Relative Atomic Mass = 106.9050916
%                                      Isotopic Composition = 0.51839
%                                      Standard Atomic Weight = 107.8682
47 109  Ag silver    0.5 -0.26112     -23/88  +0.0002436  108904755/10^6  -2155361873/16544000  -1.30280577   -99/76  -0.0001742 
%                                      Relative Atomic Mass = 108.9047553
%                                      Isotopic Composition = 0.48161
%                                      Standard Atomic Weight = 107.8682
%-------------------------------------------------------------------------------------------------------------------------------
\end{verbatim}
\end{widetext}

\section{Protons stopping in graphite\label{sec:StoppingProtons}}
From Fig.~\ref{fig:Equal-period-pp-collider-mod-mod} one sees, in right angle crossing collision geometry,
that the laboratory frame is close to, but not quite equal to the center of mass (CM) frame.  Nevertheless,
in our orthogonal collision geometry, compared to laboratory fixed target scattering measurement, scattering 
symmetries are much better preserved.  The horizontal, $x,y$, and vertical $x,z$ planes are common to both 
laboratory and CM frames, thereby preserving left/right and up/down symmetries.  Though elastically scattered 
protons are exactly collinear in the CM frame, they are more nearly orthogonal in our laboratory frame. 

To simplify discussion, we will ignore the implied slow transverse velocity in the laboratory, taking it
as the CM frame.  In this approximation, all CM scatters through $\pi/2$ lie in the same ``orthogonal'' plane, common 
to laboratory and CM frames and have azimuthal angles preserved, but polar angles are slightly distorted.  
Scattered particles have identical energies but not collinear paths.

Table~\ref{tbl:graphite-stopping} shows stopping powers and ranges for kinetic energies in the applicable range. 
With graphite density of 1.7\,g/cm$^2$, all final state protons will stop in the graphite chamber, producing accurate 
energies for most scattered protons. Precision energy determination (for example to exclude inelastic scatters) 
depends on the full stopping range.  But, since the $p,C$ polarization analyzing power falls with decreasing proton 
energy, polarimetric analyzing power is provided mainly by the left/right asymmetry of $p,C$ elastic scatters in the 
first half of their ranges, while their energies remain high.
\begin{table}[h]\scriptsize
\medskip
\centering
\begin{tabular}{|c|ccc|c|c|}             \hline
K.E. & Stopping   & Power,   & $dK/d(\rho_0l)$ & range, $\rho_0l$ & 20\,col3/col4\\
     &          & MeV cm$^2$/gm &           & gm/cm$^2$ &          \\
MeV  & electronic(e) & nuclear(n) & total(t) &          & n-prob. \\  \hline
20   & 2.331E+01  & 1.006E-02  & 2.332E+01  & 4.756E-01 & 0.00862  \\ 
40   & 1.331E+01  & 5.221E-03  & 1.331E+01  & 1.662E+00 & 0.00784  \\ 
60   & 9.642E+00  & 3.553E-03  & 9.645E+00  & 3.453E+00 & 0.00736 \\ 
80   & 7.714E+00  & 2.703E-03  & 7.717E+00  & 5.786E+00 & 0.00700 \\ 
100  & 6.518E+00  & 2.186E-03  & 6.520E+00  & 8.616E+00 & 0.00670 \\ 
120  & 5.701E+00  & 1.838E-03  & 5.703E+00  & 1.190E+01 & 0.00644 \\ 
140  & 5.107E+00  & 1.587E-03  & 5.109E+00  & 1.561E+01 & 0.00621 \\ 
160  & 4.655E+00  & 1.398E-03  & 4.656E+00  & 1.971E+01 & 0.00600 \\ 
180  & 4.299E+00  & 1.250E-03  & 4.301E+00  & 2.418E+01 & 0.00581 \\ 
200  & 4.013E+00  & 1.130E-03  & 4.014E+00  & 2.900E+01 & 0.00563 \\ \hline
sum  &            &            &            &           & 0.06761 \\
\hline
\end{tabular}
\caption{\label{tbl:graphite-stopping}Stopping power for protons stopping in graphite,
density 1.70\,gm/cm$^2$, NIST\cite{NIST-PSTAR}. Col6 gives the probability of nuclear scatter
in the approximation that nuclear energy loss (in this energy range) is always negligible compared 
to electric energy loss.  The binning error associated with quite wide kinetic energy bins
is quite small because the probabilities vary slowly.  Though the polarization detection 
efficiency $E$, is 7\% percent, the analyzing power will exceed, say 70\%, 
only for, perhaps, an order of magnitude smaller efficiency.}
\end{table}


\begin{thebibliography}{99}
\bibitem{RT-ICFA} 
R.M. Talman, \emph{Superimposed Electric/Magnetic Dipole Moment Comparator Lattice Design,} 
ICFA Beam Dynamics Newsletter \#82, Yunhai Cai, editor, 2021 JINST 16 P09006 

\bibitem{RT-CLIP-PTR} 
R. Talman, \emph{Difference of measured proton and He3 EDMs: a reduced systematics test of T-reversal invariance,}
arXiv:2205.10526v2 [physics.acc-ph], https://doi.org/10.1088/1748-0221/17/11/P11039

\bibitem{LeeWolfenstein}  
T.D.Lee and L. Wolfenstein, 
\emph{Analysis of CP-Invariant Interactions and the $K_1^0$ and $K_2^0$ system,}
Phys. Rev. {\bf 138}, 68, 1965

\bibitem{PrentkiVeltman}   
J. Prentki and M. Veltman, \emph{Possibility of CP violation in semi-strong interactions,} Phys.Letters15,88, 1965

\bibitem{Okun.1}
L. B. Okun, \emph{Remark on CP-parity,} Sov. J. Nucl. Phys., {\bf 1}, 1965  

\bibitem{Okun}  
I.Yu. Kobzarev, L.B. Okun et al., \emph{The Violation of CP Invariance,} https://doi.org/10.1070/PU1967v009n04ABEH003013

\bibitem{Gomez-time-scales}  
J. Gomez-Camacho, et al., \emph{Time scales in nuclear structure and nuclear reactions of exotic nuclei,}
Il Nuovo Cimento, 42, C, 2019

\bibitem{Derbenev}
Y.S. Derbenev et al., \emph{Siberian Snakes, Figure-8 and Spin Transparency Techniques
for High Precision Experiments with Polarized Hadron Beams in Colliders,} Symmetry, 13, 398. 
https://doi.org/10.3390/sym13030398, 2021

\bibitem{AshtekarDeLorenzoKhera}
A. Ashtekar, T. De Lorenzo, and N. Khera, \emph{Compact binary coalescences: The subtle issue of angular 
momentum}, Phys. Rev D, 101, 044005 2020



\bibitem{CYR}
CPEDM Group, \emph{Storage ring to search for electric dipole moments of charged particles
Feasibility study,} CERN Yellow Reports: Monographs, CERN-2021-003, 2021

\bibitem{Filatov}
Y. Filatov et al. \emph{Transparent spin method for spin control of
hadron beams in colliders,} Phys. Rev. Lett.,  124, 194801, 2020

\bibitem{Morse-Feshbach}
J. Blatt and V. Weisskopf, \emph{Theoretical Nuclear Physics,} Dover Publications, 1991 reprint of
Springer-Verlag, 1979, from John Wiley, 1952

\bibitem{Gamow}
G. Gamow and C. Critchfield, \emph{Theory of Atomic Nucleus and Nuclear Energy Sources,}
Scholar Select reprint of Oxford, at the Clarendon Press, 1949 

\bibitem{LeLuc-Lehar}  
C. Lechanoine-LeLuc and F. Lehar, \emph{Nucleon-nucleon elastic scattering and total cross sections,}
Rev. Mod. Phy, {\bf 65}, 1, 1993

\bibitem{MottMassey}
 N. Mott and H. Massey, \emph{The Theory of Atomic Collisions,} Oxford, at the Clarendon Press, 1965

\bibitem{NIST-background}, \emph{Background information related to the constants,} 
https://physics.nist.gov/cuu/Constants/background.html

\bibitem{NIST-isotope-abundance}, \emph{Atomic Weights and Isotopic Compositions for All Elements,} 
https://physics.nist.gov/cgi-bin/Compositions/stand\_alone.pl

\bibitem{Van-de-Hulst}, \emph{Light Scattering by Small Particles,} Dover Publications, 1981 
(originally, John Wiley and Sons, Inc. NY), 1957

\bibitem{DrellHearn}
S.D. Drell and A.C.Hearn, \emph{Exact sum rule for nucleon magnetic moments,}
Phys. Rev.   {\bf 16}, 20, 1965 

\bibitem{Gerasimov}
S.B. Gerasimov, Sov. J. Nucl. Phys. 2 430, 1966

\bibitem{Helbing}
K. Helbing, \emph{Experimental verification of the GDH sum rule,} 
arXiv:nucl-ex/0603021v3 29 Mar 2006, and Physikalisches Institut, Universitat Erlangen-N, 2018

\bibitem{BystrickyLehar.1}    
J. Bystricky, F. Lehar, and P. Winternitz, \emph{On tests of time reversal invariance in 
nucleon-nucleon scattering,} Journale de Physique, {\bf 45}, 2, pp 207-224, 1984

\bibitem{Wilkin}       
C. Wilkin, \emph{The legacy of the experimental hadron physics program at COSY,} 
Eur. Phys. J. A 53 (2017 114, 2017

\bibitem{Eversmann}      
D. Eversmann et al., \emph{New method for a continuous determination of the spin tune
in storage rings and implications for precision experiments,}
Phys. Rev. Lett. {\bf 115} 094801, 2015

\bibitem{Hempelmann}     
N. Hempelmann et al., \emph{Phase-locking the spin precession in a storage ring,} P.R.L.
119, 119401, 2017

\bibitem{Rathmann-Kolya-Slim}
F. Rathmann, N. Nikoliev, and J. Slim, \emph{Spin dynamics investigations for the electric dipole moment experiment,}
Phys. Rev. Accel. Beams 23, 024601, 2020

\bibitem{Slim-Rathmann}
J. Slim et al., \emph{First detection of collective oscillations of a stored deuteron beam
with an amplitude close to the quantum limit,} Phys. Rev. Accel. Beams, 24, 124601, 2021

\bibitem{RathmannWienFilter}
F. Rathmann, \emph{First direct hadron EDM measurement with deuterons using COSY,} 
Willy Haeberli Memorial Symposium, https://www.physics.wisc.edu/haeberli-symposium, 2022

\bibitem{RT}
R.Talman, \emph{Improving the hadron EDM upper limit using doubly-magic proton and helion beams,}
arXiv:2205.10526v1 [physics.acc-ph] 21 May, 2022

\bibitem{Talman-Nikolaev}      
R. Talman and N. N. Nikolaev, \emph{Colliding beam elastic $p,p$ and $p,d$ scattering to test $T$- and  $P$-violation,}
Snowmass~2021, Community Town Hall/86, 5 October, 2020

\bibitem{KolyaFrankPaulo}
P. Lenisa et al., \emph{Low-energy spin-physics experiments with polarized beams and targets at the COSY storage ring,}
EPJ Techniques and Instrumentation, 
https://doi.org/10.1140/epjti/s40485-019-0051-y, 2019

\bibitem{HanNambu}
M. Han and Y. Nambu, \emph{Three-Triplet Model with Double SU(3) Symmetry,} 
Phys. Rev. {\bf129}, 4B, 1965

\bibitem{Sakharov}       
A.D. Sakharov, \emph{Violation of CP invariance, C asymmetry, and baryon asymmetry of the universe,}
JETP Lett. {\bf 5}, 24-27, 1967

\bibitem{ArashMoravcsikGoldstein}     
F. Arash, M. Moravcsik, and G. Goldstein, \emph{Dynamics-independent Null, experiment 
for testing time-reversal independence,} Phys. Rev. Lett., {\bf{54}}, 2649, 1985

\bibitem{Stodolsky}      
L. Stodolsky, Nucl. Phys., \emph{Parity violation in threshold neutron scattering,} {\bf B197}, 213, 1982 

\bibitem{Ieira}       
M. Ieira, et al., \emph{A multifoil carbon polarimeter for protons between 20 and 84\,MeV,} Nuclear 
Instruments and Methods in Physics Research, {\bf A257}, 253-278, 1987

\bibitem{Plottner-A-eq=1}   
G.R. Plottner and A.D. Bocher, \emph{Absolute calibration of spin 1/2 polarization,} 
Phys. Lett. {\bf 36B}, 3, 211, 1971

\bibitem{Przewoski}     
B. von Przewoski, et al., \emph{Absolute measurement of the p +p analyzing power at 183 MeV,} 
Phys. Rev, C, {\bf 44}, 1, p44, 1991

\bibitem{Bagdasarian}    
Z. Bagdasarian et al., \emph{Measurement of the analysing power in proton-proton elastic
scattering at small angles,} Physics Letters B, {\bf 739}, 152, 2014

\bibitem{JacksonBlatt}
J.D. Jackson and J.M.Blatt, \emph{The interpretation of low energy proton-proton scattering,}
Rev. Mod. Phys. {\bf 22}, 1, 77, 1950

\bibitem{WignerEisenbud}
Feenberg and E. Wigner, \emph{Symmetry properties of Nuclear levels,}   Repts. Progr. in Phys., {\bf{8}}, 274, 1941

\bibitem{NIST-PSTAR}
National Bureau of Standards Physical Measurement Laboratory, https://physics.nist.gov

\bibitem{Stratton}
Julius Stratton, \emph{Electromagnetic Theory,} McGraw-Hill, Book Company, Inc., 1941 
\end{thebibliography}
\end{document}